\documentclass[]{aa}  
\usepackage{graphicx}
\usepackage{txfonts}
\usepackage{natbib}
\newcommand{\psrtar}{PSR\,J1846$-$0258}

\newcommand{\gtap}{\mathrel{\hbox{\rlap{\lower.55ex \hbox {$\sim$}}
                   \kern-.3em \raise.4ex \hbox{$>$}}}}
\newcommand{\ltap}{\mathrel{\hbox{\rlap{\lower.55ex \hbox {$\sim$}}
                   \kern-.3em \raise.4ex \hbox{$<$}}}}

\begin{document}
   \title{High-energy characteristics of the schizophrenic pulsar \psrtar\ in Kes 75}

   \subtitle{Multi-year RXTE and INTEGRAL observations crossing the magnetar-like outburst}

   \author{L. Kuiper\inst{1}
          \and
          W. Hermsen\inst{1,2}
          }

   \offprints{L. Kuiper}

   \institute{SRON-Netherlands Institute for Space Research, Sorbonnelaan 2, 
              3584 CA, Utrecht, The Netherlands\\
              \email{L.M.Kuiper@sron.nl}
         \and
             Astronomical Institute ``Anton Pannekoek", University of 
             Amsterdam, Kruislaan 403, 1098 SJ Amsterdam, The Netherlands\\
             \email{W.Hermsen@sron.nl}
             }

   \date{Received 23 December 2008 / Accepted 24 April 2009}

  \abstract
   {}
   {\psrtar\ is a young rotation-powered pulsar with one of the highest surface magnetic field strengths, located 
   in the centre of SN-remnant Kes-75. In June 2006 a magnetar-like outburst took place. Using multi-year RXTE and 
   INTEGRAL observations covering also the epoch of the outburst we aim to study the temporal and spectral 
   characterisitics of \psrtar\ over a broad $\sim$ 3-300 keV energy range to derive constraints on theoretical 
   scenarios aiming to explain this schizophrenic behaviour.}
   {We explored all publically available RXTE observations of \psrtar\ to generate accurate ephemerides over the 
   period January 30, 2000 - November 7, 2007. Phase folding procedures yielded pulse profiles for RXTE PCA 
   ($\sim3-30$ keV), RXTE HEXTE ($\sim15-250$ keV) and INTEGRAL ISGRI ($\sim20-300$ keV). The pulsed spectrum 
   over the full $\sim3-300$ keV energy range was derived, as well as the total spectrum (including the Pulsar Wind 
   Nebula) over the 20-300 keV band with the ISGRI. The timing, spatial and spectral analyses have been applied 
   for epochs before, during and after the magnetar-like outburst to study the evolution of the high-energy 
   characteristics.}
   {ISGRI detected \psrtar/Kes-75 before outburst during 2003-2006 with a power-law-shape spectrum over the 
   20-300 keV energy range with photon index $\Gamma$ = $1.80\pm0.06$ and energy flux (20-300 keV) of 
   $(6.62\pm 0.35)\times 10^{-11}$ erg/cm$^2$s. More than 
   90 days after the onset of the outburst, still during the decay phase, the same spectral shape was measured 
   ($\Gamma$ = $1.75_{-0.31}^{+0.27}$) with an indication for a 52\% (2.3$\sigma$) enhanced total emission, while 
   one year after the outburst the hard X-ray non-thermal emission of \psrtar/Kes-75 was found to be back to its 
   pre-outburst values. PCA monitoring of \psrtar\ before the outburst yielded phase-coherent ephemerides
   confirming the earlier derived breaking index of the spindown. During the outburst incoherent 
   solutions have been derived.
   We showed that the radiative outburst was triggered by a major spin-up glitch near MJD $53883\pm 3$ with
   a glitch size $\Delta\nu/\nu$ in the range $(2.0-4.4)\times 10^{-6}$. Using all pre-outburst observations of 
   ISGRI and HEXTE for the first time pulse profiles have been obtained up to 150 keV with a broad single 
   asymmetric pulse. The pulse shape did not vary with energy over the 2.9-150 keV energy range, nor did it 
   change during the magnetar-like outburst. 
   The time-averaged pre-outburst $\sim$ 3-300 keV pulsed spectrum measured with the PCA, HEXTE and ISGRI 
   was fitted with a power-law model with $\Gamma$ = $1.20\pm 0.01$. A fit with a curved power-law model gives an 
   improved fit. Around 150 keV the pulsed fraction approaches 100\%.
   The first 32 days of the magnetar-like outburst the 3-30 keV pulsed spectrum can be represented with two 
   power laws, a soft component with index $\Gamma_{s}$ = $2.96\pm 0.06$ and a hard component with the pre-outburst
   value $\Gamma_{h} \sim 1.2$. Above $\sim$ 9 keV all spectra during outburst are consistent with the latter single power-law shape 
   with index $\sim 1.2$. The 2-10 keV flux increased by a factor $\sim 5$ and the 10-30 keV flux increased with 
   only 35\%. After $\sim$ 120 days the soft outburst and the enhancement of the hard non-thermal component both vanish.
   }
    {The varying temporal and spectral characteristics of \psrtar\ can be explained in a scenario of a young
    high-B-field pulsar in which a major glitch triggered a sudden release of energy. Resonant cyclotron 
    upscattering could subsequently generate the decaying / cooling soft pulsed component measured during outburst 
    between 3 and 10 keV. The (variation in the) non-thermal hard X-ray component can be explained with synchrotron 
    emission in a slot-gap or outer-gap pulsar model.
    }
   \maketitle
%
\begin{table*}[t]
\caption{INTEGRAL observation characteristics for \psrtar}
\label{table:integral}
\centering
\begin{tabular}{c c c c c c c}
\hline\hline
Revs.    & Date begin & Date End   & MJD         & GTI$^1$ exposure & Eff.$^2$ exposure & \# Scw$^3$ \\
         &            &            &             &     (Ms)     &     (Ms)      &        \\
\hline
\multicolumn{7}{c}{\textit{Pre-outburst}}\\
\\
049-070  & 10-03-2003 & 13-05-2003 & 52708-52772 & 1.3737       & 0.6954        & 657\\
109-123  & 04-09-2003 & 18-10-2003 & 52886-52930 & 0.3434       & 0.1869        & 180\\
172-233  & 11-03-2004 & 10-09-2004 & 53075-53258 & 1.1416       & 0.5047        & 402\\
236-250  & 18-09-2004 & 01-11-2004 & 53266-53310 & 0.6488       & 0.3142        & 343\\
251-315  & 02-11-2004 & 14-05-2005 & 53311-53504 & 0.6681       & 0.3945        & 315\\
345-379  & 10-08-2005 & 21-11-2005 & 53592-53695 & 1.0405       & 0.4418        & 347\\
407-441  & 12-02-2006 & 26-05-2006 & 53778-53881 & 0.7707       & 0.4604        & 355\\
\hline
049-441  & 10-03-2003 & 26-05-2006 & 52708-53881 & 5.9868       & 2.9979        &2599\\
\\
\multicolumn{7}{c}{\textit{Outburst/Post-outburst}}\\
\\
474-501  & 31-08-2006 & 19-11-2006 & 53978-54058 & 0.3721       & 0.2135        & 148\\
537-561  & 07-03-2007 & 20-05-2007 & 54166-54241 & 0.3908       & 0.1561        & 130\\
592-603  & 19-08-2007 & 23-09-2007 & 54331-54366 & 0.6011       & 0.2587        & 168\\
\hline
474-603  & 31-08-2006 & 23-09-2007 & 53978-54366 & 1.3640       & 0.6283        & 446\\
\\
\multicolumn{7}{c}{\textit{All observations}}\\
\\
049-603  & 10-03-2003 & 23-09-2007 & 52708-54366 & 7.3508       & 3.6262        &3045\\
\hline
\multicolumn{7}{l}{$^1$ Total Good-Time-Interval exposure of the used observations}\\
\multicolumn{7}{l}{$^2$ Effective exposure on \psrtar\ corrected for off-axis sensitivity reduction}\\
\multicolumn{7}{l}{$^3$ Number of used Science Windows, see text}\\
\end{tabular}
\end{table*}

\section{Introduction}

\psrtar\ was discovered  in a timing analysis of X-ray data from RXTE and ASCA by \citet{gotthelf2000}. It is a 
relatively slow rotation-powered pulsar, $P \sim 324$ ms, and has one of the highest surface magnetic 
field strengths, $B \sim 4.9 \times  10^{13}$ G, assuming standard magnetic dipole breaking, just above the 
quantum critical field strength $B_{cr}=m_ec^3/e\hbar$ of $4.413 \times 10^{13}$ G. Furthermore, it has the smallest 
characteristic age, $\tau \sim 723$ y, of all known pulsars and a spin-down luminosity ($\dot{E}_{sd}$) of $8.2\times 10^{36}$ erg/s. 
At radio frequencies it remains undetected \citep{archibald08}.
\psrtar\ is located at the centre of Supernova Remnant (SNR) Kes 75 \citep[G29.7-0.3;][]{kesteven1968} and 
was resolved in a Chandra observation as a bright X-ray source surrounded by a diffuse pulsar wind nebula \citep[PWN;][]{helfand2003}. 
The distance of SNR Kes 75 and \psrtar\ has been under discussion for a long time. The most recent estimates are 
by \citet{leahy2008}, 5.1 to 7.5 kpc, and \citet{su2008}, $\sim 10.6$ kpc. For a discussion on earlier estimates we refer
to their papers. In this work we adopt a distance of 10 kpc, and write the distance to Kes 75 as $d=d_{10}\times 10$ kpc.

Based on analysis of more than five years of RXTE monitoring data it appeared that \psrtar\/
behaves as a very stable rotator with a breaking index of $n=2.65\pm 0.01$ \citep[$\dot\nu\propto\nu^{n}$;][]{livingstone06}.
INTEGRAL detected hard X-ray point-source emission from the PWN plus \psrtar\ up to $\sim$ 200 keV \citep{mcbride2008}, and 
HESS discovered the PWN in the TeV band \citep{djannati07}.

Recently, \citet{kumar08} reported the detection of a dramatic brightening of the pulsar during June, 7-12, 2006 in Chandra 
observations of Kes 75. The pulsar's spectrum (pulsed plus DC emission) softened considerably from a power-law index 
$\Gamma \sim 1.32$ measured in 2000 to $1.97$ during the outburst in 2006. \citet{gavriil08} used RXTE observations to
show that this prompt radiative event exhibited a recovery from the pulsed-flux enhancement which can be modeled
as an exponential decay with 1/{\slshape e} time constant of $55.5 \pm 5.7$ day and a total 2-60 keV energy release of 
$(3.8-4.8)\times 10^{41}$ erg, adopting a distance of 6 kpc \citep{leahy2008} and assuming isotropic emission.
For the revised distance estimate of $\sim10$ kpc the released energy becomes $(1.1-1.3)\times 10^{42}d_{10}^2$.
Such an energy release is of similar magnitude as has been detected from a few Anomalous X-ray Pulsars (AXPs).
The outburst was accompanied by an unprecedented change in timing behaviour.
Furthermore, \citet{gavriil08} also discovered short ($<0.1$ s) magnetar-like bursts during the outburst, 4 at the 
beginning and 1 near the end.
This otherwise very stably behaving rotating neutron star, both temporally and spectrally, switched suddenly to 
unpredictable  magnetar-like behaviour. This is the first time that such dual characteristics have been seen. 
Therefore, the magnetar-like outburst of \psrtar\ warrants a more detailed study to obtain tighter constraints 
for modelling this temporary magnetar-like manifestation of a rotation-powered pulsar.

The main motivation of this paper is to study the X-ray spectral and timing characteristics before, during and
after the outburst including for the first time the hard X-ray regime 10-300 keV.
In Section 2 we introduce the instruments used in this work, the PCA and HEXTE aboard NASA's Rossi X-ray Timing 
Explorer (RXTE) and IBIS-ISGRI aboard ESA's INTErnational Gamma-Ray Astrophysics Laboratory (INTEGRAL). In Section 3 
we present an INTEGRAL deep-exposure sky map of the Scutum region (20-70 keV) revealing \psrtar\,/Kes 75 surrounded by 
several near-by sources, and derive for energies above 20 keV three spectra for the total emission, PWN and \psrtar, 
namely, pre-outburst, during outburst and post-outburst. Section 4 deals with the timing studies. First, the high 
statistics in the PCA data are exploited to derive phase-coherent ephemerides for \psrtar, pre-outburst, for part of 
the outburst and post-outburst, as well as incoherent solutions during the outburst, revealing a major spin-up glitch 
at the start of the outburst. These (in)coherent timing solutions are subsequently used to derive pulse profiles over 
the energy windows covered by the PCA, HEXTE and ISGRI for the pre-outburst observations, and using just the PCA for 
the outburst period. In Section 5 we derive the spectrum of the pre-outburst total-pulsed emission using PCA, HEXTE 
and ISGRI data and compare this with the total, PWN and \psrtar, spectra measured with Chandra and INTEGRAL-ISGRI. 
Furthermore, we use PCA data to study the evolution of the spectrum of the pulsed signal pre-outburst, during the 
outburst and post outburst. Finally, the results are summarized in Section 6 and discussed in Section 7.
\section{Instruments}

\subsection{RXTE}

In this study extensive use is made of data from monitoring observations
of \psrtar\/ with the two non-imaging X-ray instruments aboard RXTE, the Proportional Counter Array 
(PCA; 2-60 keV) and the High Energy X-ray Timing Experiment (HEXTE; 15-250 keV). The PCA 
\citep{jahoda96} consists of five collimated xenon proportional 
counter units (PCUs) with a total effective area of $\sim 6500$ cm$^2$ over a $\sim 1\degr$ 
(FWHM) field of view. Each PCU has a front Propane anti-coincidence layer and three Xenon 
layers which provide the basic scientific data, and is sensitive to photons with energies in 
the range 2-60 keV. The energy resolution is about 18\% at 6 keV. All data used in this work have
been collected from observations in {\tt GoodXenon} or {\tt GoodXenonwithPropane} mode allowing 
high time resolution ($0.9\mu$s) studies in 256 spectral channels.

The HEXTE instrument \citep{rothschild98} consists of two independent detector 
clusters, each containing four Na(Tl)/ CsI(Na) scintillation
detectors. The HEXTE detectors are mechanically collimated to a $\sim 1\degr$ (FWHM) 
field of view and cover the 15-250 keV energy range with an energy resolution of 
$\sim$ 15\% at 60 keV. The collecting area is 1400 cm$^2$ taking into account the 
loss of the spectral capabilities of one of the detectors. The best time 
resolution of the tagged events is $7.6\mu$s. In its default operation mode the 
field of view of each cluster is switched on and off source to provide instantaneous 
background measurements.
Due to the co-alignment of HEXTE and the PCA, they simultaneously observe the 
same field of view.

\begin{figure*}[t]
\centering
\includegraphics[angle=90,width=19cm,height=9cm,clip=]{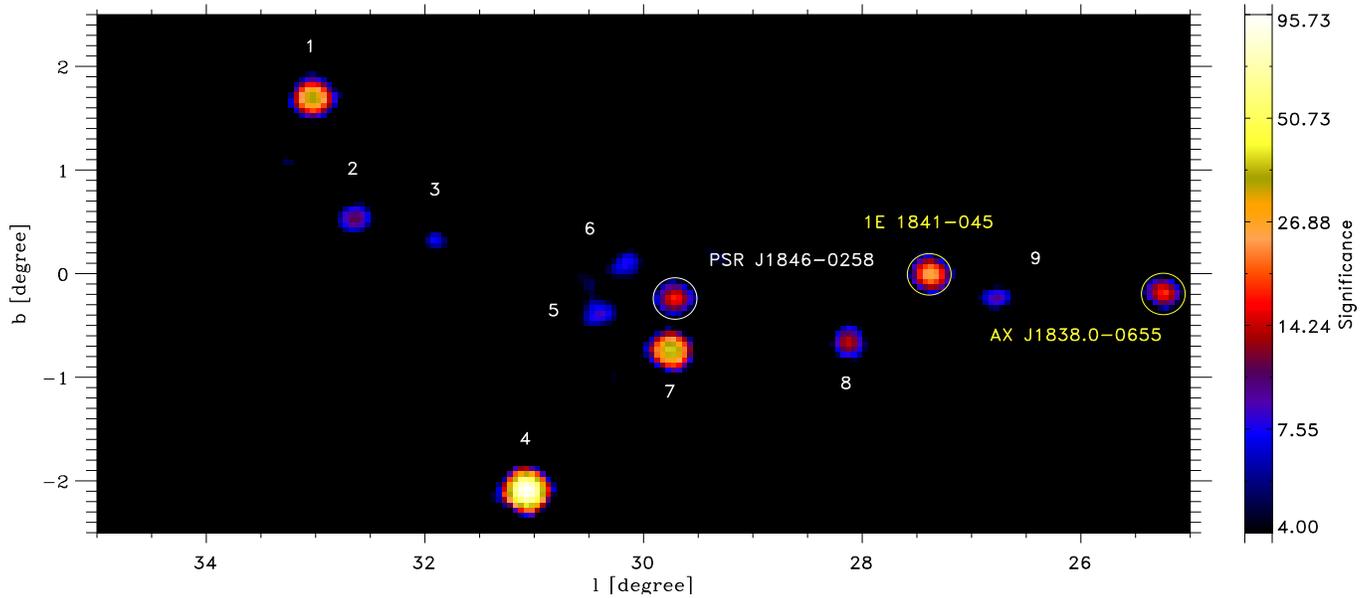}
\caption{A 20-70 keV image in Galactic coordinates of the Scutum region centered on (l,b)=(30,0) combining IBIS ISGRI data from INTEGRAL revolutions 49 up to and including 603 (March 10, 2003 - Sept. 23, 2007). The total GTI exposure amounts 7.35 Ms for this mosaic image, while the effective exposure on \psrtar\ equals 3.63 Ms. Only source features with significances above $4\sigma$ are shown. The white circle indicates \psrtar\//Kes 75 . The yellow encircled sources are AXP 1E 1841-045 \citep[see][]{kuiper04,kuiper06} and the rotation-powered pulsar AX J1838.0-0655 \citep[][]{gotthelf08,kuiper08}. The other sources indicated by numbers are: [1] - GS 1843+009,[2] - IGR J18490-0000, [3] - IGR J18485-0047, [4] - XTE J1855-026, [5] - 3A 1845-024, [6] - IGR J18462-0223, [7] - IGR J18483-0311, [8] - IGR J18450-0435 and [9] - AX J1841.0-0535}
\label{isgri_image}
\end{figure*}

\subsection{INTEGRAL}

The INTEGRAL spacecraft \citep{winkler03}, launched 17 October 2002, carries two main 
$\gamma$-ray instruments: a high-angular-resolution imager IBIS \citep{ubertini03} and
a high-energy-resolution spectrometer SPI \citep{vedrenne03}. Both instruments are equiped with  
coded aperture masks enabling image reconstruction in the hard X-ray/soft $\gamma$-ray band.

Driven by sensitivity considerations, we used only data from the INTEGRAL Soft Gamma-Ray Imager 
ISGRI \citep{lebrun03}, the upper detector layer of IBIS, sensitive to photons with energies in 
the range $\sim$20 keV -- 1 MeV. 
With an angular resolution of about $12\arcmin$ and a source location accuracy of better 
than $1\arcmin$ (for a $>10\sigma$ source) this instrument is able to locate and separate 
high-energy sources in crowded fields within its $19\degr \times 19\degr$ 
field of view (50\% partially coded) with an unprecedented sensitivity ($\sim$ 960 cm$^2$ at 50 keV).
Its energy resolution of about 7\% at 100 keV is amply sufficient to determine the (continuum) spectral 
properties of hard X-ray sources in the $\sim$ 20 - 300 keV energy band.

The timing accuracy of the ISGRI time stamps recorded on board is about $61\mu$s. The time 
alignment between INTEGRAL and RXTE is better than $\sim 25\mu$s \citep[see e.g.][]{kuiper03,falanga05}.
Given the fact that the accuracy of the RXTE clock in absolute time is about $2\mu$s \citep{rots04},
this implies that the INTEGRAL absolute timing is better than $\sim 27 \mu$s. 
Data from regular INTEGRAL Crab monitoring observations show that the clock behaviour is stable, allowing 
timing studies of weak pulsars by accumulating data taken over many years.

In its default operation mode INTEGRAL observes the sky in a dither pattern 
with $2\degr$ steps, which could be rectangular e.g. a $5 \times 5$ dither pattern 
with 25 grid points, or hexagonal with 7 grid points (target in the middle). Typical integration
times for each grid point (pointing/sub-observation) are in the range 1800 - 
3600 seconds. This strategy drives the structure of the INTEGRAL data archive which is 
organised in so-called science windows (Scw) per INTEGRAL orbital revolution (lasting for 
about 3 days) containing the data from all instruments for a given pointing. 

\section{ISGRI Imaging and Total Spectra}

\subsection{Deep ISGRI Sky Map of Scutum Region}

In this work we exploited the good imaging capability of ISGRI, which allowed us to separate \psrtar/Kes 75
from nearby sources in this crowded region in the first Galactic quadrant. ISGRI sky mosaics (a combination of deconvolved images of single science windows) were produced in 10 logarithmically binned
energy bands covering the 20-300 keV energy window. We used the imaging software tools \citep{goldwurm03} of the Offline 
Scientific Analysis (OSA) package version 5.1 distributed by the INTEGRAL Science Data Centre 
\citep[ISDC; see e.g.][]{courvoisier03}.
The end product of this analysis provides the count rate, its variance, exposure and significance for all 10 energy bands over the (deconvolved) mosaiced sky field. 

In the observation selection we only accepted science windows with instrument pointings within $14\fdg5$ from 
the \psrtar\/ position. This ensures that (a part of) the detector plane is illuminated by the target. The resulting list is further screened on erratic count rate variations, indicative for particle-induced effects due to Earth-radiation-belt passages or solar-flare activities, by inspecting visually the count rate in 20-30 keV band versus time. Science windows showing erratic count-rate variations are excluded for further analysis.
In the event-selection process we only use events with rise times between 7 and 90 \citep[see][for definition]{lebrun03}, detected in non-noisy ISGRI detector pixels.
 
The INTEGRAL observations used in this work are listed in Table \ref{table:integral}. 
Combining all these publicly available data from INTEGRAL revolutions 49 up to and including 603 we obtained a total exposure of 7.35 Ms (Good-Time-Intervals, GTI). Fig. \ref{isgri_image} shows for the 20-70 keV range (weighted sum of 5 energy bands) the significance image of the field surrounding \psrtar/Kes 75. In this image we adopted a detection significance threshold of $4\sigma$. Clearly visible at the center of the image with high significance ($17.4\sigma$) is the emission from \psrtar\ and its PWN (not resolved). Other interesting high-energy sources are also indicated in this $10\degr\times5\degr$ image of the Scutum field, particularly Anomalous
X-ray Pulsar 1E 1841-045, embedded in SNR Kes 73, and AX J1838.0-0655, which harbours a young energetic pulsar as found recently by
\citet{gotthelf08}. The identifications of the other sources are given in the figure caption.

The total spectrum of \psrtar\ and Kes 75 can be constructed from the 10 count-rate and variance maps by extracting the (dead-time corrected) rates and uncertainties at the location of \psrtar.
These values are normalized to the count rates measured for the total (nebula and pulsar) emission from the Crab in similar energy bands using Crab calibration observations during INTEGRAL revolutions 102 and 103. From the ratios and the photon spectrum of the total emission from the Crab, we can derive the total high-energy photon spectrum of the PWN plus \psrtar\ (pulsed and any unpulsed point source component) without detailed knowledge of the ISGRI energy response. For the total Crab photon spectrum we use the broken-power-law spectrum derived by \citet{jourdain08} based on INTEGRAL-SPI observations of the Crab at energies between 23 and 1000 keV. The latest Crab cross calibrations between SPI and IBIS-ISGRI provided consistent results.

\subsection{Total pre-outburst spectrum in 20-300 keV band}

Following the above procedure, the pre-outburst (Revs. 49-441, see Table \ref{table:integral}) IBIS-ISGRI spectrum was derived and 
can be fitted with a power-law model over the 20-300 keV energy range yielding a photon index $\Gamma$ = $1.80\pm0.06$ and an energy 
flux (20-300 keV) of $(6.62\pm 0.35)\times 10^{-11}$ erg/cm$^2$s ($\chi^2_{\nu}=0.878$ for 10-2 degrees of freedom; errors 
are $1\sigma$, throughout this paper). This index is a slightly harder, but consistent with the value $2.0\pm0.2$ obtained by 
\citet{mcbride2008}, who used less data. 
The 20-100 keV energy flux derived in this work $(3.47\pm 0.19)\times 10^{-11}$ erg/cm$^2$s, however, is about 20\% higher than the value, $(2.9_{-0.1}^{+0.2})\times 10^{-11}$ erg/cm$^2$s, 
obtained by \citet{mcbride2008}. This discrepancy can be fully attributed to their use of obsolete energy response matrices for 
IBIS ISGRI in the OSA 5.1 environment, which yield inconsistent spectral results for the total Crab spectrum. Our ISGRI flux measurements 
for the total pre-outburst emission between 20 and 300 keV from the PWN plus \psrtar\/ are shown in Fig. \ref{he_spectrum} as purple data 
points with the best power-law fit superposed, and will be discussed later together with the spectrum of the pulsed emission.
\label{sect_tot_spc}

\subsection{Total spectrum (20-300 keV) during outburst decay}

Unfortunately, INTEGRAL was not pointing to the Scutum region during the early phase of the outburst.
The first exposure started on MJD 53978 (Rev. 474), more than 90 days after the onset of the outburst, with the source
still showing enhanced pulsed flux levels about 40\% higher than the pre-outburst level (see e.g. top panel of Fig. \ref{outburst_freq_evol}).
A spectrum of the total emission from \psrtar/Kes 75 from INTEGRAL observations taken between revolutions 474 and 501 (0.2135 Ms effective on-source exposure time; more than 95\% of the exposure has been accumulated during Revs. 474 \& 475) could adequately be described by a power-law with photon index $\Gamma$ = $1.75_{-0.31}^{+0.27}$
and energy flux (20-100 keV) of $(5.27_{-0.7}^{+0.8})\times 10^{-11}$ erg/cm$^2$s. The spectral index is fully consistent with that of the pre-outburst spectrum, but the total flux is about 52\% higher than the pre-outburst 
flux value, in line with the increased pulsed flux level shown in PCA data at the same epoch at lower energies. The increase, however, represents only a $\sim 2.3\sigma$ effect 
relative to its pre-outburst value. Therefore, we can not claim a significant increase of the total 20-100 keV flux, but only an indication for
enhanced total emission in the 20-100 keV band. 
\label{sect_tot_outspc}

\subsection{Total post-outburst spectrum in 20-300 keV band}
\label{sect_tot_postspc}
About one year after the onset of the outburst INTEGRAL observed \psrtar/Kes 75 for an effective exposure of 0.415 Ms during MJD 54166-54366 (INTEGRAL Revs. 537-603, see Table \ref{table:integral}).
Its derived total 20-300 keV spectrum could be described by a power-law model with $\Gamma$ = $1.90_{-0.35}^{+0.33}$ and a 20-100 keV energy flux of
$(3.64_{-0.54}^{+0.59})\times 10^{-11}$ erg/cm$^2$s, fully consistent with the pre-outburst measurement.

\begin{table*}[t]
\caption{Phase-coherent ephemerides for \psrtar\ as derived from RXTE PCA (monitoring) data.}
\label{eph_table}
\begin{center}
\begin{tabular}{lccclllcc}
\hline
Entry &  Start &  End  &   t$_0$, Epoch   & \multicolumn{1}{c}{$\nu$}   & \multicolumn{1}{c}{$\dot\nu$}               & \multicolumn{1}{c}{$\ddot\nu$}                  & $\Phi_{0}$
  & Validity range\\
 \#   &  [MJD] & [MJD] &     [MJD,TDB]    & \multicolumn{1}{c}{[Hz]}    & \multicolumn{1}{c}{$\times 10^{-11}$ Hz/s}  & \multicolumn{1}{c}{$\times 10^{-21}$ Hz/s$^2$}  &           
  &  (days)   \\
\hline\hline
\\
\multicolumn{8}{c}{\textit{Pre-Outburst}}\\
\\
0         & 51286 & 51290 & 51286.0     & 3.08273789(23)   & -6.98(15)              &   0.0 (fixed)              & 0.6850  & 4\\
\vspace{-2mm}\\
1$^1$     & 51574 & 52237 & 51574.0     & 3.0810613509(18) & -6.73176(1)            &   3.821(4)                 & 0.9825  & 663\\
2$^2$     & 52369 & 52837 & 52523.0     & 3.0755528807(5)  & -6.707486(9)           &   3.86(3)                  & 0.6501  & 468\\
3         & 52915 & 53148 & 52915.0     & 3.0732836450(59) & -6.6963(1)             &   4.98(13)                 & 0.1971  & 233\\
4         & 53030 & 53465 & 53030.0     & 3.0726185382(34) & -6.69068(3)            &   3.93(2)                  & 0.5030  & 435\\
5$^3$     & 53464 & 53880 & 53464.0     & 3.0701124555(45) & -6.67599(4)            &   3.78(3)                  & 0.8985  & 416\\
\\
\multicolumn{8}{c}{\textit{Outburst-B}}\\
\\
6         & 53978 & 54033 & 53997.0     & 3.0669883665(67) & -6.8688(7)             &   161(15)                  & 0.3092  & 55\\
\\
\multicolumn{8}{c}{\textit{Post-Outburst; coherence fully caught-up}}\\
\\
7         & 54126 & 54229 & 54126.0     & 3.066215110(19)  & -6.8726(8)             &   202(2)                   & 0.2658  & 103\\
8         & 54237 & 54340 & 54237.0     & 3.065565121(15)  & -6.7297(7)             &   15(2)                    & 0.9120  & 103\\
9         & 54340 & 54411 & 54340.0     & 3.064967085(26)  & -6.707(2)              &   49(6)                    & 0.8056  &  71\\
\hline
\multicolumn{9}{l}{$^1$ Small glitch reported by \citet{livingstone06} near MJD $52210\pm 10$}\\
\multicolumn{9}{l}{$^2$ Phase coherence lost over 78 d data gap starting at MJD 52837 \citep[see also][]{livingstone06} }\\
\multicolumn{9}{l}{$^3$ Phase coherence lost after MJD 53880 (May 24, 2006) during the magnetar-like outburst \citep[][]{gavriil08} }\\
\end{tabular}
\end{center}
\end{table*}

\section{Timing}

\subsection{Timing solutions: ephemerides}
\label{sect_tm_sol}

RXTE observed \psrtar\, for the first time in the period April 18-21, 1999 (MJD 51286-51290) for about 39 ks.
The observation was split in 11 short observations (sub-observations) of duration ranging from 2.5 to 19.9 ks.
We barycentered the PCA event arrival times using the Chandra X-ray observatory (CXO) sub-arcsecond position 
of \psrtar, $(\alpha,\delta)=(18^{\hbox{\scriptsize h}}46^{\hbox{\scriptsize m}}24\fs94,-02\degr58\arcmin30\,\farcs1)$ 
for epoch J2000 \citep{helfand2003}, which corresponds to (l,b)=(29.712015,-0.240245) in Galactic 
coordinates.
In each of these sub-observations a coherent pulsed signal at a rate of $\sim 3.0827$ Hz \citep[see also][]{gotthelf2000} 
could be detected in the barycentered time series. We used a $Z_1^2$-test \citep{buccheri1983} search in a small, typically 5
independent Fourier steps ($\Delta\nu_{\hbox{\scriptsize IFS}}=1/\tau$, in which $\tau$ represents the time span of the data period), window around the predicted pulse frequency.
The restricted search yielded for each sub-observation a best estimate of the rotation rate at the gravity point
of the sub-observation. Because the $Z_1^2$-test is distributed as a $\chi^2$ for $2\times 1$ degrees of freedom a $1\sigma$ error 
estimate on this optimum value can easily be derived by determining the intersection points of the measured $Z_1^2$-test 
distribution near the optimum with the value of $Z_{\hbox{\scriptsize 1,max}}^2-2.296$. The set of optimum pulse frequencies and $1\sigma$
uncertainties versus time can be used to obtain an incoherent timing solution $\nu_{inc}(t)$ for the rotation behaviour of 
the pulsar.
Phase folding the arrival times of each sub-observation on the appropriate optimum frequency results in pulse profiles for each sub-observation.
In order to derive the mutual phase relation of the pulse maxima we use cross-correlation techniques to obtain 
a high-statistics pulse-profile template. This template, based on 11 sub-observations, was subsequently used in the time of 
arrival (TOA) determination procedure, described below. 

The accuracy of the timing parameters\footnote{In this work we limited ourselves to a maximum of three timing parameters or less in case of small time spans.} $(\nu,\dot\nu,\ddot\nu)$ can be improved considerably relative to the incoherent set by 
maintaining pulse-phase coherence over the entire time interval of the considered set of sub-observations.
This requires an accurate determination of pulse arrival times for each sub-observation in the time interval considered.
These times are obtained by cross-correlating the pulse profiles of each sub-observation with the high-statistics template. The global 
maximum in the correlation diagram determines the time shift ($\Delta\Phi_{\hbox{\scriptsize max}}/\nu$) to be applied to the chosen time zero point of the sub-observation (normally the gravity point) to align it with the template.
The global maximum of the correlation function is derived by fitting a truncated Fourier series (typically 5 harmonics are used) to 
the measured correlation function, evaluated at discrete step points, in order to suppress fluctuations. 
The latter could be significant in case of weak pulsed signals.
An error estimate $\delta\Phi$ on $\Delta\Phi_{\hbox{\scriptsize max}}$ has been obtained by bootstrapping both the pulse-profile 
of the active sub-observation and that of the template assuming an underlying Gaussian distribution. 
Typically 1000 resamplings are used in subsequent correlation analysis. The $1\sigma$ error on $\Delta\Phi_{\hbox{\scriptsize max}}$ 
corresponds to the width of the distribution of the correlation maxima. Thus, we obtained a set of $n$ pulse-arrival times with 
corresponding error estimates, $(t_{\hbox{\scriptsize p}}^n,\Delta t_{\hbox{\scriptsize p}}^n)$.
Next, the arrival times were folded upon the incoherent or any other reasonable timing solution to obtain the residuals 
$$\Delta\Phi_k=\nu_{inc}\cdot(t_{\hbox{\scriptsize p}}^k-t_{\hbox{\scriptsize p}}^0) + \frac{1}{2}\dot\nu_{inc}\cdot(t_{\hbox{\scriptsize p}}^k-t_{\hbox{\scriptsize p}}^0)^2 + \frac{1}{6}\ddot\nu_{inc}\cdot(t_{\hbox{\scriptsize p}}^k-t_{\hbox{\scriptsize p}}^0)^3$$

Finally, we determined the increments $\Delta\nu,\Delta\dot\nu$ and $\Delta\ddot\nu$ which minimize the $\chi^2$ statistics
$\sum_{k=1}^{n-1}(\frac{\Delta\Phi_k}{\delta\Phi_k})^2$ taking into account the statistical weights of the individual points.
This approach yields a phase-coherent (or phase-connected) timing solution $(\nu_c,\dot\nu_c,\ddot\nu_c)$ describing every revolution of
the pulsar over the time span for which the solution is valid.

As of January 30, 2000 RXTE monitors \psrtar\ regularly, and from observations performed till November 7, 2007 (MJD 51574-54411) 
we derived phase coherent timing solutions (ephemerides) when possible. Table \ref{eph_table} lists the current set of phase-coherent 
ephemerides, as derived in this work using the method described above, including the one (entry \# 0) covering the four days of the 
very first RXTE observations of \psrtar.

\citet{livingstone06} presented equivalent timing solutions for the range 
MJD 51574-53578, covering about 5.5 year of X-ray timing data. Our timing results cross the magnetar-like radiative outburst event
that occurred between May 24-31, 2006 \citep[see e.g.][]{gavriil08,kumar08}. Entries 1 till 5 of Table \ref{eph_table} cover
the so-called pre-outburst period, which is characterized by very stable and highly predictable timing behaviour allowing an
accurate measurement of the breaking index of the pulsar spin-down of $2.65\pm0.01$ \citep{livingstone06}. Our timing results for the
pre-outburst period (MJD 51574-53880) are fully compatible with those obtained by \citet{livingstone06} for a slightly smaller 
time span (MJD 51574-53578). It demonstrates that \psrtar\/ continued to slow-down very predictably from MJD 53578 till MJD 53880, the
date of the last sub-observation before the magnetar-like outburst. 

\subsection{A small and a major spin-up glitch}
\label{sect_tm_glitch}

We also confirm the loss of phase-coherence beyond MJD 52237, attributed by \citet{livingstone06} to a small glitch of size $\Delta\nu/\nu=2.5(2)\times 10^{-9}$ and $\Delta\dot\nu/\dot\nu\sim9.3(1)\times10^{-4}$ near MJD $52210\pm 10$.
Note, however, that the TOA for the sub-observation on MJD 52237 can still perfectly be predicted by the timing parameters of entry \# 1 
of Table \ref{eph_table}, indicating that the small glitch very likely occured beyond MJD 52237, but before MJD 52253, the next 
sub-observation, thus later than the glitch epoch of $52210\pm 10$ reported by \citet{livingstone06}.

\begin{figure}[t]
  \hspace{0.3cm}\includegraphics[width=8.7cm,height=6.5cm]{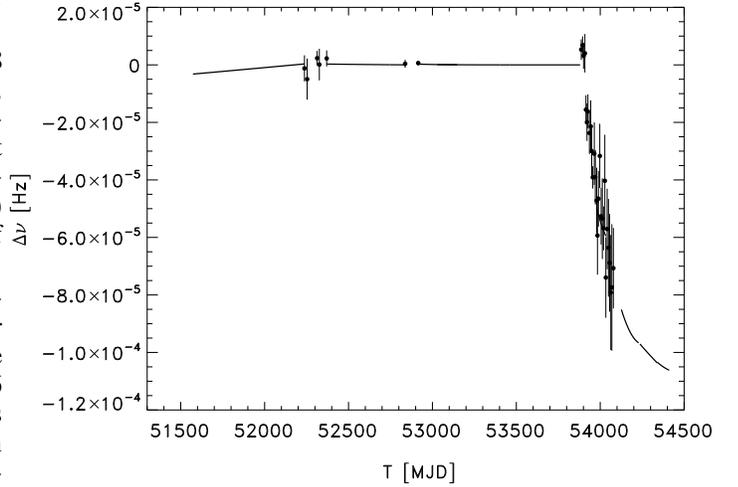}
  \caption{The rotation behaviour of \psrtar\ from Jan. 30, 2000 to Nov. 7, 2007 as determined from RXTE monitoring observations
         relative to the pre-outburst timing model (entry \# 5 of Table \ref{eph_table}).
         Note the dramatic change during the radiative outburst and its subsequent decay (e-folding time scale $\sim 55$ days).
         The presence of a huge spin-up glitch near MJD $53883 \pm 3$ is clearly visible.
         Phase coherence is caught-up again as of MJD 54126.Lines: phase-coherent timing models. Data  points with 1-$\sigma$ errors: incoherent solutions.}
\label{rotation_behaviour}
\end{figure}

As of the onset, beyond MJD 53880, of the radiative outburst till MJD 54126 (January 26, 2007) phase coherence is completely lost due
to highly increased timing noise \citep[see also][]{gavriil08}. The sparse sampling in this period and the high timing noise exclude
the construction of a phase coherent solution during the outburst and its subsequent decay except for a small period of 55 days from 
MJD 53978 till 54033 (entry \#6 of Table \ref{eph_table}). Phase coherence is caught-up again on MJD 54126 (entries \#'s 7 and 8 of Table
\ref{eph_table}). Note the very high value of $\ddot\nu$ and the increased $|\dot\nu|$ values of entries \#'s 6 and 7.
All timing information both coherent and incoherent, if the former is lacking, is shown in Fig. \ref{rotation_behaviour}. In this 
figure the frequencies (coherent, solid lines; incoherent, data points) are relative to the timing model valid just before the radiative
outburst (entry \# 5 of Table \ref{eph_table}).

From Fig. \ref{rotation_behaviour} it is clear that the radiative outburst was accompanied by a huge spin-up glitch near MJD $53883\pm 3$
as suspected already by \citet{gavriil08}, but now clearly demonstrated. 
Depending on the number of sub-observations used in the construction of an incoherent timing model since the onset of the outburst 
we obtain a glitch size $\Delta\nu/\nu$ in the range $(2.0-4.4)\times 10^{-6}$. This fractional-frequency-jump size ranks in the 
top of the glitch-size distribution for rotation-powered pulsars, but is typical for AXP glitches \citep{dib08}, although the 
number of observed AXP glitches is still too small to securely establish the underlying distribution.

\begin{figure}
 \centering
 \includegraphics[height=7cm,width=8cm,bb=35 188 512 632,clip=]{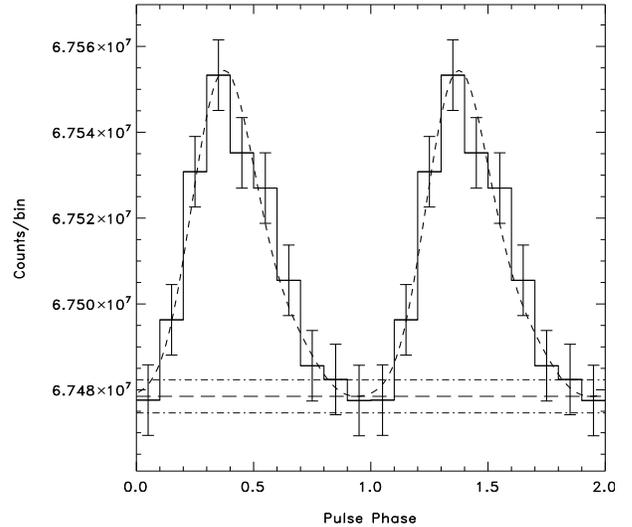}
\caption{The pre-outburst (Revs. 49-441) pulse profile of \psrtar\ as observed by IBIS ISGRI in the 20-150 keV energy band with a detection significance of $9.6\sigma$. Error bars are $1\sigma$ on measured counts. Superposed is the best fit model composed of a PCA-template profile for the $\sim$ 2.9-8.3 keV band (dashed line) and a DC-level (long dashes; horizontal line) with its $1\sigma$-error estimates (see Sect. \ref{sect_pes_anm}). There is no significant difference in shape.}
\label{isgri_pulse_profile}
\end{figure}

\begin{figure*}[t]
\centering
\includegraphics[angle=90,width=15cm,height=9.8cm,bb=90 145 470 660,clip=]{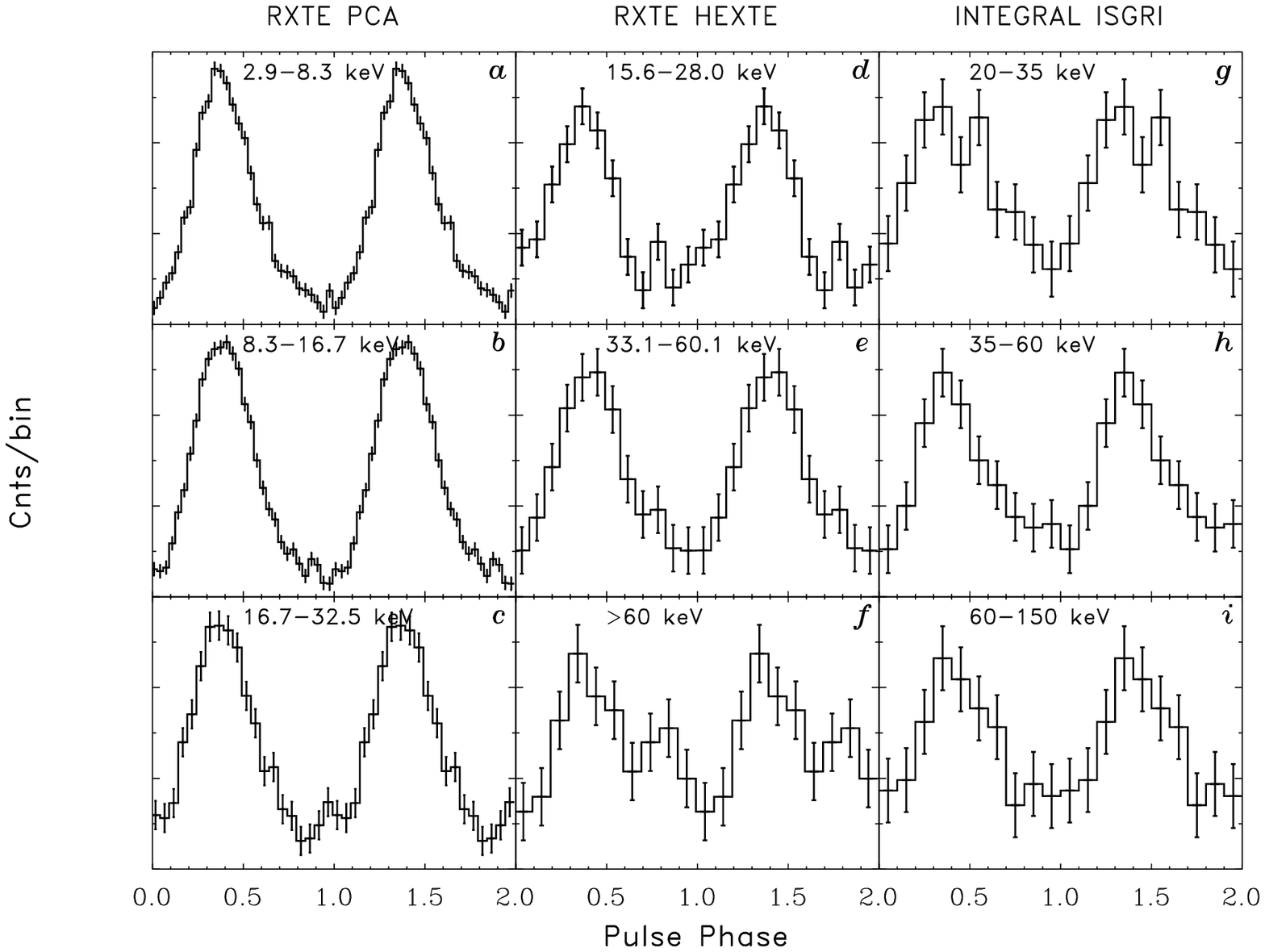}
\caption{Collage of pulse profile of \psrtar\ using pre-outburst data from RXTE PCA (3-35 keV: a-c), HEXTE (15-250 keV: d-f) and INTEGRAL ISGRI (20-300 keV: g-i).
The pulse signal is detected for the first time up to $\sim 150$ keV making \psrtar\/ the fifth pulsar showing up at soft $\gamma$-rays. The pulse profile shape is energy independent: a broad single-peak asymmetric pulse. Error bars are $1\sigma$ on measured counts.}
\label{profile_collage}
\end{figure*}

\subsection{Pre-outburst pulse profiles: INTEGRAL ISGRI, RXTE PCA and HEXTE}
\label{sect_tm_pp}

The timing analysis of INTEGRAL ISGRI data consists of selecting, screening and finally phase folding barycentered time series with
appropriate ephemerides. The observation and event selection procedures are identical to those applied for the imaging studies of ISGRI data. In order to reduce in the timing analysis the backgound level consisting of non-source counts, only detector pixels are considered which have been illuminated by the source for at least 25\%.
In this work, we further screen the data on short duration (lasting less than a second) bursts, particularly those originating from the soft gamma-ray repeater SGR 1806-20, which pollute the genuine pulsed signal from \psrtar.
Then, the on-board-registered event-time stamps of the selected events are corrected for known instrumental (fixed), ground station and general time delays in the on-board-time versus Terrestrial-Time (TT or TDT) correlation \citep[see e.g.][]{walter03}.
The resulting event times in TT are barycentered (using the JPL DE200 solar system ephemeris) adopting the CXO position of \psrtar\ and 
the instantaneous INTEGRAL orbit information. These barycentered events (in TDB time scale) are finally folded upon the appropriate timing model composed of $\nu,\dot\nu,\ddot\nu$ and the epoch $t_0$, as obtained in the phase-coherence analysis of RXTE PCA data (see Table \ref{eph_table}). The TDB-time to pulse-phase conversion taking into account consistent profile alignment by subtracting 
$\Phi_0$ is provided by the following formula:
$$\Phi(t)=\nu\cdot (t-t_0) + \frac{1}{2}\dot\nu\cdot (t-t_0)^2+\frac{1}{6}\ddot\nu\cdot (t-t_0)^3 - \Phi_0$$
Thus, we produced pulse-phase distributions for differential energy bands of width 1 keV between 15 and 300 keV.
For the ``stable-timing-behaviour'' period before the onset of the magnetar-like outburst we obtained a $9.6\sigma$ signal (Z$_1^2$-test) 
in the 20-150 keV band (see Fig. \ref{isgri_pulse_profile}) combining all timing data from INTEGRAL observations performed between
revolutions 49 and 441 (see Table \ref{table:integral}). The total Good-Time-Interval (GTI) exposure of the 2599 Scw's included in this combination amounts $\sim 6$ Ms, which translates in an effective on-source exposure of $\sim 3$ Ms. The right hand panels g, h and i of Fig.
\ref{profile_collage} show the ISGRI pulse profiles in the three differential energy bands, 20-35 keV, 35-60 keV and 60-150 keV.
The pulsed-signal detection significances are $5.6\sigma, 7.0\sigma$ and $4.2\sigma$, respectively, making
 \psrtar\/ the fifth pulsar detected in the soft $\gamma$-ray band, after the Crab, Vela, PSR B1509-58 and PSR B0540-69 pulsars.

We wish to compare the ISGRI pulse-phase distributions with the time-averaged HEXTE and PCA pulse profiles from all RXTE observations performed before
the magnetar-like outburst. The timing analysis of HEXTE data, covering an energy window comparable to ISGRI, is equivalent to the approach followed by \citet{kuiper06} (see 
Sect. 3.1 of that paper). We collected 344.122 ks and 335.770 ks exposure, both corrected for dead time and reduction in efficiency
due to off-axis observations, for HEXTE cluster A and B, respectively. Phase folding HEXTE barycentered-time series with
the ephemerides derived from the simultaneously taken PCA data (see Table \ref{eph_table}) yields the high-energy phase distributions from HEXTE. Differential HEXTE profiles are shown in panels d-f of Fig. \ref{profile_collage} for the 15.6-28 keV, 33.1-60.1 keV and $>60.1$ keV energy bands\footnote{Spectral data from the 28-33.1 keV HEXTE band have been ignored because of the presence of a huge background line due to the activation of iodine.}, respectively. The corresponding significances are $11.0\sigma, 8.7\sigma$ and $4.3\sigma$. The ISGRI and HEXTE profile shapes are fully consistent. Finally, we also included in Fig. \ref{profile_collage} the PCA pulse profiles from pre-outburst observations for energies between $\sim2.9$ and $\sim32.5$ keV. As can be seen, the morphology of the pulse profile for energies above $\sim 2.9$ keV appears to be energy independent: a broad single-peak asymmetric pulse with a somewhat steeper rise than fall.
\begin{figure}
  \vspace{-0.25cm}
  \resizebox{0.975\hsize}{!}{\includegraphics[]{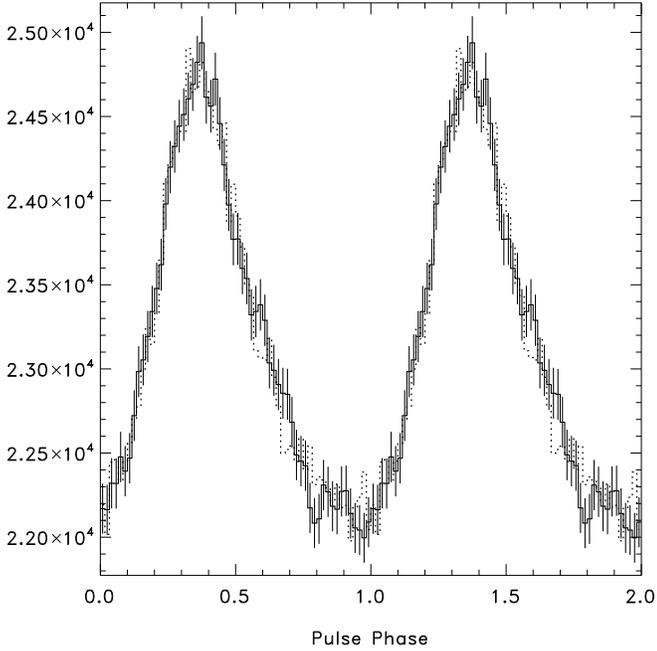}}
  \vspace{0.25cm}
\caption{Comparison of the pulse-profiles accumulated in the PCA PHA band 7-19 ($\sim 2.9-8.3$ keV) during the early phase of the outburst
(Outburst A1+A2, see Table \ref{table:pca_temp_spec}, solid histogram with $1\sigma$ error bars) and from the pre-outburst observations (high-statistics dotted histogram). No significant shape change is observed, see text.
The Y-axis specifies the number of counts per bin.}
\label{profile_comparison}
\end{figure}
\subsection{Pulse-profile morphology during the outburst}
\label{sect_tm_pmorph}
To investigate whether the radiative outburst, most pronounced at soft X-rays, was accompanied by a pulse-morphology change we compared the high-statistics pre-outburst PCA profile with the profile from RXTE-PCA observations during the early phase of the outburst.
For the latter we used data collected between MJD 53894 and 53970 (Outburst A1 + A2, see Table \ref{table:pca_temp_spec} and also Fig. \ref{outburst_freq_evol}), omitting those observations in which the reported five short (0.1 s) magnetar-like bursts occurred. We selected only events from PHA channels 7 to 19, roughly corresponding to 2.9-8.3 keV. Because phase coherence was lost after the onset of the outburst, we generated the ``outburst" profile by stacking the properly shifted profiles from individual observations applying a cross-correlation analysis which yields the required phase shifts \citep[see eg. Section 3.1 of][for a similar analysis]{kuiper04}. Fig. \ref{profile_comparison} shows the superposed aligned pre-outburst and outburst profiles. We fitted the outburst profile in terms of a constant and the shape of the pre-outburst profile with a free scale.
The absolute values and signs of the deviations from the best fit were used in a combination of two independent statistical tests, a Pearson $\chi^2$ test and a run test. The combined probability of these tests is 3.2\%, i.e. both profiles are different at a $2.15\sigma$ significance level, indicating that there is {\em no} significant change in shape between the pre-outburst and outburst profile.

\section{Pulsed Emission Spectra}

\subsection{Analysis methods}

\label{sect_pes_anm}

From the pulse-phase distributions, $F(\Phi,E_{\hbox{\scriptsize PHA}})$, derived for RXTE PCA, HEXTE and INTEGRAL IBIS ISGRI, we
extracted pulsed excess counts by fitting two different model functions to the measured pulse-phase distributions in selected energy bands: 
\begin{itemize}
   \item[1)] a truncated Fourier series with $N-1$ harmonics,\\ 
   $N(\Phi)=a_0+\sum_{k=1}^{N}a_k\cos(2\pi k\Phi)+b_k\sin(2\pi k\Phi)$\\
   \item[2)] a PCA-template model, $N(\Phi)=a+b\times T(\Phi)$ 
\end{itemize}
For model 1, a number of $N=3$ was sufficient to adequately describe the measured profiles. The template function $T(\Phi)$ used for
model 2 is based on a high-statistics PCA pulse profile for energies between $\sim 2.9-8.3$ keV.

The fit function minimum plus its error are subsequently used to determine the number of pulsed excess counts i.e. the number of counts above this minimum (=unpulsed or DC level) along with an error estimate. For pulse profiles with a strong pulsed signal both methods yielded consistent results, however, for weak pulsed signals method 1 overestimates the number of pulsed excess counts slightly (a small positive bias). Therefore, we have used the template fit model subsequently, assuming thus that the genuine underlying pulse profile did not vary with energy which is a very reasonable approximation (see e.g. Fig. \ref{isgri_pulse_profile}).

For the PCA we constructed time-averaged energy response matrices for each PCU separately taking into account the different (screened) exposure times of the involved PCU's during the time period of interest. For this purpose we used the {\it ftools version 6.4} programs {\it pcarsp} and {\it addrmf}. To convert PHA channels to measured energy values, $E_{\hbox{\scriptsize PHA}}$, for PCU combined/stacked products we also generated a weighted PCU-combined energy response matrix.

For HEXTE we employed cluster A and B energy-response matrices separately, taking into account the different screened on-source exposure times and the reduction in efficiency in case of off-axis observations. The on-source exposure times for both clusters have been corrected for considerable dead-time effects.

We assume simple power-law models, $F_{\gamma}= K\cdot (E_{\gamma}/E_0)^{-\Gamma}$ with $\Gamma$ the photon-index and $K$ the normalization in ph/cm$^2$s keV at the pivot energy $E_0$, for the underlying photon spectra. In case of RXTE PCA\footnote{An absorbing Hydrogen column of N$_{\hbox{\scriptsize H}}=3.96\times 10^{22}$ cm$^{-2}$ has been assumed for the PCA spectral analysis \citep[see][]{helfand2003} and in all 
PCA spectra shown in this paper the interstellar absorption has been modeled out.} and HEXTE these models have been fitted in a forward folding procedure using appropriate response matrices to obtain the optimum spectral parameters, $K$ and $\Gamma$, and the reconstructed spectral flux points from the observed pulsed count rates. 

A different approach has been used for IBIS ISGRI: For each energy band we scaled the derived pulsed excess counts by the number of pulsed excess counts extracted for the Crab pulsar in exactly the same measured energy window. These ratios were subsequently multiplied by the pulsed photon spectrum of the Crab as derived from deep HEXTE observations \citep[see Sect. 3.4 of][for details]{kuiper06} taking into account the different effective on-source exposures for both \psrtar\/ and the Crab.
\begin{figure}
  \resizebox{\hsize}{!}{\includegraphics[]{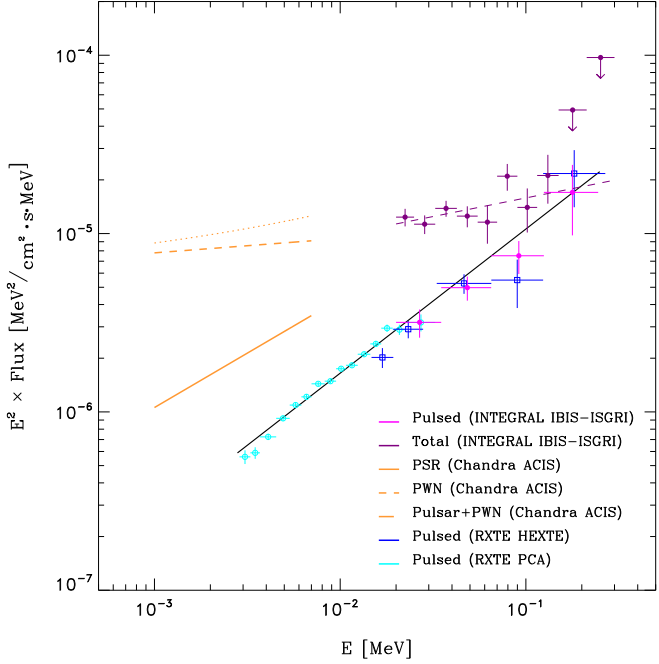}}
\caption{Total and pulsed (unabsorbed) high-energy spectrum ($\sim 1-300$ keV) of Kes75 / \psrtar\/ from pre-outburst 
RXTE PCA and HEXTE (both pulsed) and INTEGRAL IBIS ISGRI (pulsed and total) observations in an $ E^{2}F$ representation.
The spectral coverage is extended towards 1 keV using the Chandra spectral-fit results (1-7 keV; unabsorbed) obtained by \citet{helfand2003} for the total emission spectrum of the pulsar and the PWN.
Note that both the PWN contribution and DC emission from \psrtar\/ drop to undetectable levels at energies
near 150 keV: the total emission is consistent with being 100\% pulsed above $\sim 150$ keV. Error bars on data points are $1\sigma$.}
\vspace{-0.45cm}
\label{he_spectrum}
\end{figure}

\subsection{Pulsed pre-outburst 3-300 keV spectrum}
\label{sect_pes_ppre}
The reconstructed pulsed emission spectra from pre-outburst RXTE and INTEGRAL observations are shown in Fig. \ref{he_spectrum} as aqua coloured data points for the PCA ($\sim$3-30 keV), blue data points for HEXTE (15-250 keV) and magenta coloured symbols for INTEGRAL ISGRI (20-300 keV). The individual PCA, HEXTE and ISGRI spectra are mutually consistent in overlapping energy bands. 

A single power-law model fit to the PCA data points ($\sim$ 3-30 keV) alone gave an optimum photon index $\Gamma$ of $1.161\pm 0.004$.
However, the obtained reduced $\chi_{\hbox{\scriptsize{r}},13}^2$ of 2.38 for 13 degrees of freedom is poor, and fitting a power-law with an energy dependent index, a so-called ``curved power-law model" $F_{\gamma}= K\cdot (E_{\gamma}/E_0)^{-\tilde{\Gamma}-\delta\cdot\ln(E_{\gamma}/E_0)}$, provided a significant improvement of $3.4\sigma$ over the simple power-law model, mainly due to better describing the flux measurements below 4.5 keV.

In order to obtain an accurate description of the pulsed spectrum over two decades in energy we also fitted a power-law and curved power-law model over the full $\sim$ 3-300 keV energy band to the combined PCA, HEXTE and ISGRI pulsed flux measurements. The best-fit power-law model, shown as a solid black line in Fig. \ref{he_spectrum}, has a photon index $\Gamma$ of $1.20\pm 0.01$, slightly softer than the fit to just the fluxes below 30 keV. The energy fluxes of the pulsed emission (unabsorbed) in the 2-10 keV and 20-100 keV bands are $(2.38\pm0.02)\times 10^{-12}$ erg/cm$^2$s and $(15.2\pm0.14)\times 10^{-12}$ erg/cm$^2$s, respectively.
Assuming a 1 steradian beam size for the pulsed emission and a distance of 10 kpc the luminosities in the 2-10 and 20-100 keV
bands are $L_{X}^{2-10}=2.27(2)\times 10^{33}\cdot \eta \cdot d_{10}^2$ erg/s and $L_{X}^{20-100}=1.45(1)\times 10^{34}\cdot \eta \cdot d_{10}^2$ erg/s ($\eta$ is the beamsize), respectively. These numbers convert into the following efficiencies ($L_X/\dot{E}_{sd}$)\,, $\epsilon_{X}^{2-10} = 0.027\% \cdot \eta \cdot d_{10}^2$ and $\epsilon_{X}^{20-100} = 0.18\% \cdot \eta\cdot d_{10}^2$\,, respectively.
The fit with a curved power-law model was again better ($4.9\sigma$ improvement over the power-law model; it is shown in Fig. \ref{spc_comparison_rpp_axp}) for the following best-fit parameter values: $K=(8.40\pm0.12)\times 10^{-6}$ 
ph/cm$^2$s keV, $\tilde{\Gamma}=1.295\pm0.016$ and $\delta=0.096\pm0.014$ for a pivot energy of 17.6209 keV.

\subsection{Pulsed fractions}
\label{sect_pes_pf}
In Fig. \ref{he_spectrum} the total (i.e. \psrtar\/ pulsed plus unpulsed/DC and PWN contribution) pre-outburst ISGRI spectrum (see Sect. \ref{sect_tot_spc}) is presented as purple data points across the 20-300 keV band as well as the best fit power-law model to these measurements (purple dashed line). Using the energy fluxes of pulsed and total emission in the 20-100 keV band we could derive a lower-limit on the pulsed fraction of 44\% in the 20-100 keV band. Moreover, it is evident that near 150 keV the pulsed emission becomes consistent with the total emission i.e. 100\% pulsation. As a result, both, the spectra of the DC-emission from the pulsar and of the PWN must bend down between 20 and 100 keV to significantly lower levels.

To extend the spectral coverage to lower energies we included in Fig. \ref{he_spectrum} the spectral model fits obtained from Chandra data in the soft X-ray band (1-7 keV) by \citet{helfand2003} \citep[see for recent analyses][]{kumar08,ng08} for the total emission from \psrtar\ ($\Gamma = 1.39\pm 0.04$; pulsed plus unpulsed/DC; solid orange line) and the PWN ($\Gamma = 1.92\pm 0.04$; dashed orange line), separately. We also show the sum of these two models as a dotted orange line representing the total emission from \psrtar/PWN in the 1-7 keV band, to be compared with the total emission as measured by ISGRI above 20 keV.
Dividing the pulsed emission estimated in the 0.5-10 keV band through extrapolation to lower energies of the $\sim$ 3-300 keV pulsed spectrum, $(2.98\pm0.03)\times 10^{-12}$ erg/cm$^2$s, by the total point-source emission $9.50 \times 10^{-12}$ erg/cm$^2$s \citep[][]{helfand2003} from \psrtar\ in the Chandra 0.5-10 keV band we obtained an average pulsed fraction of 31\%.
Therefore, comparing this value with the lower-limit of 44\% derived for the 20-100 keV band, the pulsed fraction is steadily growing with energy to become within the large errors consistent with $\sim 100\%$ near 150 keV. This is primarily an indication that the DC spectra have to bend down.

\begin{figure}
  \centering
  \includegraphics[height=12cm,width=9cm,bb=5 60 555 715]{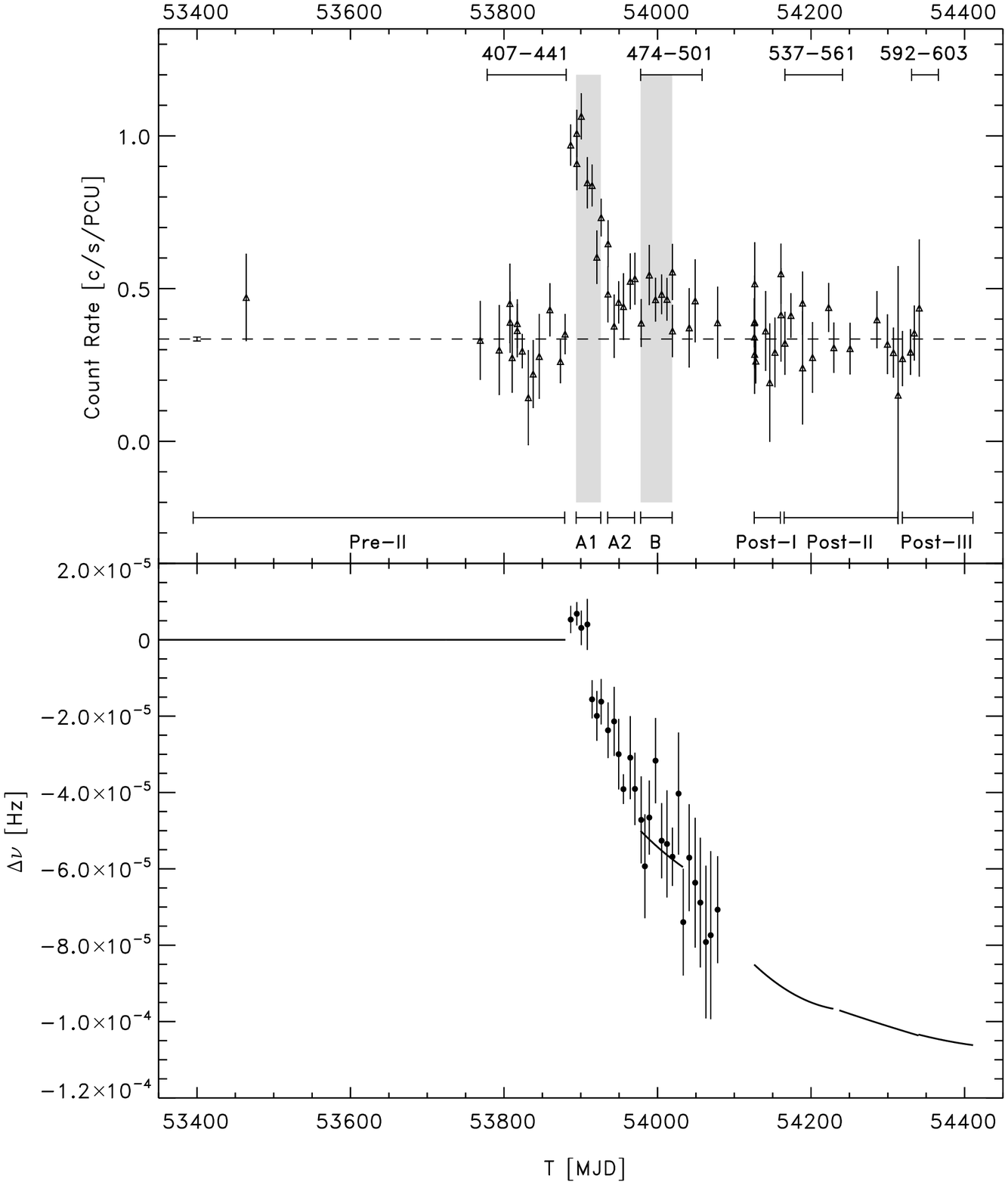}
\caption{Pulsed-flux lightcurve (PCA 2-60 keV) around the outburst (top panel) adapted from \citet{gavriil08}. The pre-outburst averaged pulsed-flux level is shown by the horizontal dashed line along with its tiny $1\sigma$ error (plotted at MJD 53400). The coverage of the INTEGRAL observations near the outburst period are indicated with revolution-number intervals in the upper part of the top panel. In the lower part of the top panel are shown the time segments used in the spectral-evolution study of the pulsed signal using 3-30 keV PCA data. The lower panel displays the pulse frequency evolution across the outburst period (zoom-in of Fig. \ref{rotation_behaviour}). The phase-coherent solution in the time interval of 55 days around MJD 54000 is indicated. Errors are $1\sigma$.}
\label{outburst_freq_evol}
\end{figure}

\begin{table*}[t]
\caption{Time intervals used in the spectral analysis of the RXTE PCA data.}
\label{table:pca_temp_spec}
\centering
\begin{tabular}{c c c c c c c l}
\hline\hline
Obs. id.    & MJD        &    \multicolumn{5}{c}{Screened PCU exposure [ks]}               & Name\\
            & Begin/End  &     0       &      1       &      2      &     3      &     4   &     \\
\hline
\vspace{-3mm} \\
40140-01-01-01  & 51286  &  940.624    &   217.960    &   940.608   &  814.520   &  350.200 & Pre-outburst I  \\
90071-01-10-00  & 53340  &             &              &             &            &          &                 \\
\vspace{-3mm} \\ \cline{1-1} \vspace{-3mm} \\
90071-01-12-00  & 53395  &  203.712    &    33.768    &   207.944   &  107.176   &   13.120 & Pre-outburst II \\
92012-01-12-00  & 53879  &             &              &             &            &          &                 \\
\vspace{-3mm} \\ \hline \vspace{-3mm} \\
92012-01-14-00  & 53894  &   20.064    &     4.880    &    24.008   &  13.712    &    2.832 & Outburst A1$^{1}$\\
92012-01-19-00  & 53926  &             &              &             &            &          &                 \\
\vspace{-3mm} \\ \cline{1-1} \vspace{-3mm} \\
92012-01-20-00  & 53935  &   11.096    &     9.616    &    20.896   &   6.720    &    2.992 & Outburst A2$^{2}$\\
92012-01-25-00  & 53970  &             &              &             &            &          &                 \\
\vspace{-3mm} \\ \cline{1-1} \vspace{-3mm} \\
92012-01-26-00  & 53978  &   23.032    &     8.232    &    31.176   &  10.176    &    9.664 & Outburst B      \\
92012-01-32-00  & 54019  &             &              &             &            &          &                 \\
\vspace{-3mm} \\ \hline \vspace{-3mm} \\
90071-01-18-00  & 54126  &   32.096    &     3.952    &    39.632   &   7.568    &   13.064 & Post-outburst I \\
91071-01-16-00  & 54160  &             &              &             &            &          &                 \\
\vspace{-3mm} \\ \cline{1-1} \vspace{-3mm} \\
91071-01-17-00  & 54165  &   78.048    &    11.330    &    91.360   &  22.192    &   32.440 & Post-outburst II \\
93010-01-05-00  & 54313  &             &              &             &            &          &                 \\
\vspace{-3mm} \\ \cline{1-1} \vspace{-3mm} \\
93010-01-06-00  & 54319  &   52.688    &    12.496    &    72.888   &  18.392    &   18.792 & Post-outburst III \\
93010-01-19-00  & 54411  &             &              &             &            &          &                 \\
\hline
\hline
\multicolumn{8}{l}{$^{1}$Obs.id. 92012-01-13-00 on MJD 53886 has been excluded because of the occurence of 4 SGR-like bursts}\\
\multicolumn{8}{l}{$^{2}$Obs.id. 92012-01-21-00 on MJD 53943 has been excluded because of the occurence of 1 SGR-like burst}\\
\end{tabular}
\end{table*}

\begin{figure*}
  \sidecaption
  \includegraphics[width=12cm]{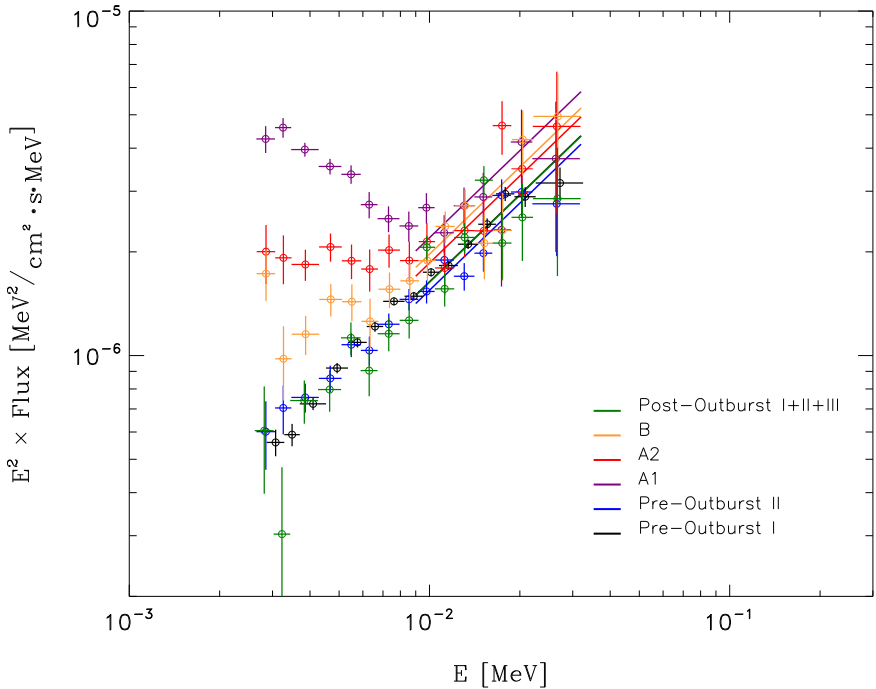}
  \caption{Evolution of the (unabsorbed) pulsed spectrum ($\sim$3-30 keV) of \psrtar\/ as derived from RXTE PCA data.
           Note the dramatic softening of the spectrum during the early phase of the outburst (purple data 
           points) and the gradual return to its pre-outburst spectrum (black data points). Also indicated 
           for each time segment are the power-law model fits obtained from $>9$ keV data. Data points are given with $1\sigma$ errors.}
  \label{spectral_variations}
\end{figure*}

\subsection{Spectral evolution (3-30 keV) before, during and after the outburst}
\label{sect_pes_spcevol}
The X-ray spectral index of the total emission from \psrtar\ (pulsed plus DC emission) measured with Chandra between 1 and 7 keV changed from a pre-burst value in 2000 of $\Gamma \sim 1.35$ \citep{helfand2003,kumar08} to a value of $\Gamma \sim 1.93$ \citep{kumar08,gavriil08} during the magnetar-like outburst in 2006. In order to better characterize the spectral evolution we took advantage of the high-statistics 3-30 keV PCA data to derive spectra for the pulsed emission before, during and after the outburst.

Fig. \ref{outburst_freq_evol} indicates in the top panel the selected time segments for the PCA analysis in comparison with the count-rate lightcurve (2-60 keV) of \psrtar\ published by \citet{gavriil08}: Segment Pre-II covers an interval just before the outburst; segments A1, A2 and B cover the reported outburst; Post-I to Post-III are time intervals after the outburst for which the count-rate lightcurve indicates that \psrtar\ returned to its pre-burst flux level. Table \ref{table:pca_temp_spec} shows the details of the PCA time segments including the identifiers of the first and last sub-observation along with the corresponding start and end time in MJD. Also, the screened exposure time per PCU is listed together with the name of the time segment.

The bottom panel of Fig. \ref{outburst_freq_evol} shows that we derived for each of these time segments coherent or incoherent timing solutions, allowing us to produce pulse profiles and subsequently spectra of the pulsed emission for each segment.
Comparing the upper and bottom panel of Fig. \ref{outburst_freq_evol}, it is clear that the major glitch triggered the radiative outburst.

\begin{figure}
  \centering
  \includegraphics[height=8cm,width=8.25cm]{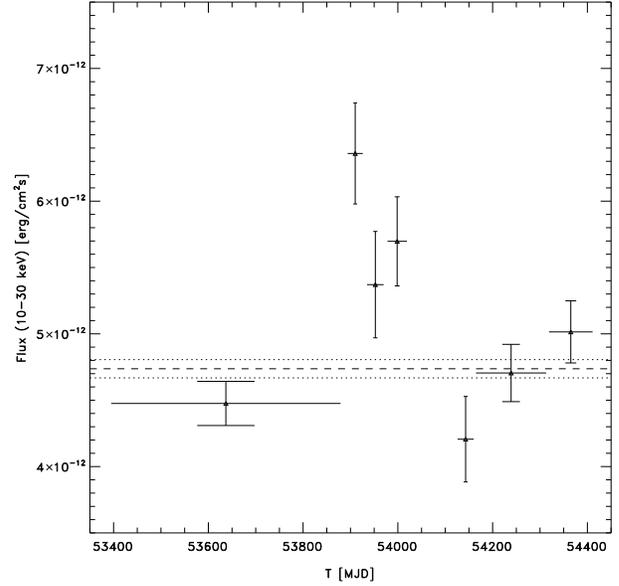}
\caption{Evolution of the pulsed flux of \psrtar\/ in the 10-30 keV band as derived from a power-law fit to RXTE PCA $>9$ keV spectra 
(photon index fixed to its pre-outburst value of 1.16). Error bars are 1$\sigma$. The average flux level in the Pre-outburst I time segment 
(duration $\sim$ 5.6 yrs) is indicated with the broken line together with its $\pm1\sigma$ uncertainty (dotted lines).}
\label{tp_outburst_gt9kev}
\end{figure}

For each PCA observation-time segment we had sufficient counting statistics to produce pulsed spectra ($\sim$3-30 keV).
Fig. \ref{spectral_variations} shows the (unabsorbed) pulsed spectra  for all time segments listed in Table \ref{table:pca_temp_spec}, including the multi-year-average spectrum of Pre-outburst I, but combining the three post-outburst spectra. The latter three spectra are combined, because 
these are statistically identical and the individual flux values could not be distinguished when plotted in this figure. 
It is immediately evident that the pulsed spectrum evolved smoothly in flux and shape during the burst. 
A dramatically softer pulsed spectrum was measured directly after the glitch (time interval A1, purple data points). Obviously, a strong 
soft component was added to the pre-burst hard ($\Gamma\sim 1.16$) non-thermal spectrum, totally dominating the pulsed emission below $\sim 10$ keV. During time segment A1 the pulsed (energy) flux for energies 2-10 keV increased by a factor $\sim$ 5.
This soft component quickly reduced in intensity during the outburst (A2, red; B, orange data points), and is not detected anymore in the post-outburst spectra which are within statistics identical to the Pre-outburst I and II spectra. In Fig. \ref{spectral_variations} one can see that the time variability is less pronounced for energies above $\sim 9$ keV. In fact, we verified that for all selected time intervals the 
spectra above 9 keV are fully compatible with a single power-law model shape with the pre-outburst index $\Gamma = 1.16$. 
In Fig. \ref{spectral_variations} these power-law model fits, all with a fixed photon index $\Gamma$ of $1.16$, are superposed for each time segment. The corresponding pulsed fluxes (10-30 keV) are shown in Fig. \ref{tp_outburst_gt9kev}, including also the individual data points for the three post-outburst time segments. We see, that also in this case the flux enhancement was maximal directly after the glitch in time segment A1, but amounts only $\sim$ 35\%, which represents a 4.3$\sigma$ enhancement.

It is evident that during the outburst the spectral shape over the 3-30 keV band varied. A single power-law model fit for time segment A1
gave a poor fit ($\Gamma$ = $2.46\pm 0.01$ with a reduced $\chi^2_{r}=1.69$ for 13 degrees of freedom). A model fit with two (free) power-laws rendered a $3.2\sigma$ improvement. An excellent fit is obtained with a soft power-law index $\Gamma_{s}$ = $2.96\pm 0.06$ along with a hard photon index of $\Gamma_{h} = 1.16$, fixed at the pre-outburst value. The pulsed spectra in time segments A2 and B have larger statistical 
errors and can be fitted with single power-laws. However, when fitted with two power-laws, these are also consistent with the sum of two spectral components, a soft one with index $\Gamma_{s} = 3.27\pm 0.27$ (segment A2) and $\Gamma_{s} = 4.74\pm 0.71$ (segment B), along with $\Gamma_{h} = 1.16$. Note the apparent gradual softening of the soft component below $\sim$ 10 keV during the decay of the outburst, indicative for cooling.

\begin{figure}
  \centering
  \includegraphics[height=8cm,width=8.25cm]{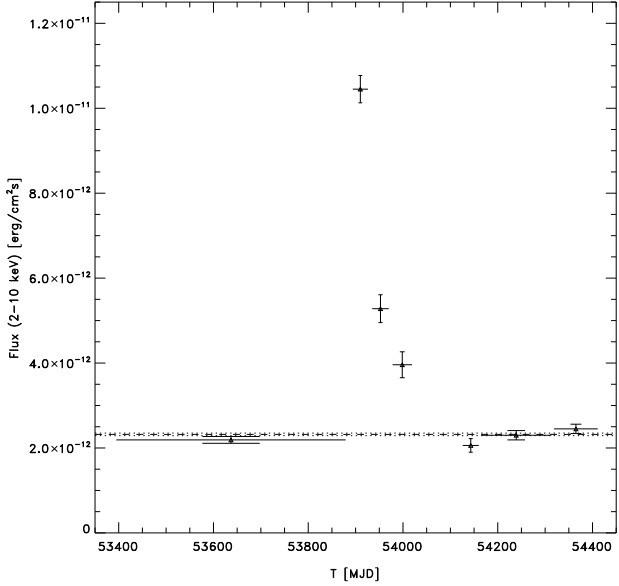}
\caption{Evolution of the pulsed flux of \psrtar\/ in the 2-10 keV band as derived from RXTE PCA data. Error bars are 1$\sigma$.
         The average non-thermal flux level in the Pre-outburst I time segment (duration $\sim$ 5.6 yrs) is indicated with its
         $\pm 1\sigma$ uncertainty like in Fig. \ref{tp_outburst_gt9kev}, but the levels practically coincide.}
   \label{excess_ph_flux_lt10}
\end{figure}

In Fig. \ref{excess_ph_flux_lt10} we present the variation of the measured pulsed fluxes below 10 keV. These are derived using the above two-power-law fits in the outburst time segments, and the single power-law fit (index 1.16) for pre- and post-outburst.
The decay of the excess photon flux measured above the non-thermal pre-outburst flux level (indicated in Fig. \ref{excess_ph_flux_lt10}) is consistent with the exponential decay with 1/{\slshape e} time constant of $\sim 55.5$ day reported by \citet{gavriil08}.
 

\section{Summary}

In this paper we have presented detailed high-energy characteristics of the enigmatic
\psrtar\ and its PWN, using the complementary spatial, spectral and timing capabilities of
RXTE PCA and HEXTE and INTEGRAL's IBIS/ISGRI. Particularly important is to see how the
characteristics compare before, during and after the unique magnetar-like outburst in order
to recognize constraints which are important for the interpretation of the transient 
magnetar-like behaviour of \psrtar.

\subsection{Total spectra above 20 keV of \psrtar/Kes 75, before, during and after outburst}

1 -- Using all available INTEGRAL data taken with ISGRI during 2003-2006 before the magnetar-like
outburst in June 2006, adding up to 3.0 Ms effective on-source exposure, the time-averaged 
20-300 keV spectrum for the total emission of \psrtar/Kes 75 can be represented by a power-law with photon index
$\Gamma$ = $1.80\pm0.06$ and an energy flux (20-300 keV) of $(6.62\pm 0.35)\times 10^{-11}$ erg/cm$^2$s
({\slshape {Sect. \ref{sect_tot_spc}}}\/).

\noindent
2 -- ISGRI observed \psrtar/Kes 75 unfortunately only towards the end of the outburst for 214 ks. No change in spectral shape was observed ($\Gamma$ = $1.75_{-0.31}^{+0.27}$),  but there is an indication at the 
$2.3\sigma$ level for an enhanced flux level (by $\sim 52\%$) during outburst decay ({\slshape {Sect. \ref{sect_tot_outspc}}}\/).

\noindent
3 -- About one year after the onset of the outburst, the measured (415 ks) total post-outburst spectrum 
in the 20-300 keV band is within 1$\sigma$ errors of $\sim$ 15\% consistent in shape and flux with the pre-outburst spectrum 
({\slshape {Sect. \ref{sect_tot_postspc}}}\/).

\subsection{Ephemerides of \psrtar\ before, during and after outburst; glitches}

1 -- Exploiting the multi-year monitoring observations of RXTE and the high count rate of the PCA
we derived phase-coherent ephemerides for the 6.5 years before the outburst, consistent with 
the results reported by \citet{livingstone06}. \psrtar\ was manifesting itself as a stable young 
high B-field pulsar with a breaking index of the spin-down of $2.65 \pm 0.01$ ({\slshape{Sect. \ref{sect_tm_sol}, Table \ref{eph_table}}}\/).

\noindent
2 -- During the outburst phase coherence was lost \citep[see also][]{gavriil08}, except for a short period
of 55 days from MJD 53978 till 54033 in which we found a phase-coherent solution. Nevertheless, incoherent 
solutions have been derived over the full duration of the outburst. Phase coherence was caught-up again after 
the burst on MJD 54126, when the PCA (2-60 keV) pulsed flux was back to its average pre-burst value ({\slshape{Sect. \ref{sect_tm_glitch}, Table \ref{eph_table}, Fig. \ref{outburst_freq_evol}}}\/).

\noindent
3 -- We showed that the onset of the radiative outburst was accompanied by a huge spin-up glitch near MJD $53883\pm 3$ with
a glitch size $\Delta\nu/\nu$ in the range $(2.0-4.4)\times 10^{-6}$. This fractional-frequency-jump size ranks in the 
top of the glitch-size distribution for rotation-powered pulsars. Furthermore, we confirm the occurrence of a small glitch 
near MJD $52210\pm 10$ \citep{livingstone06}, however, for a slightly different epoch somewhere between MJD 52237 and 52253 
({\slshape {Sect. \ref{sect_tm_glitch}, Fig. \ref{outburst_freq_evol}}}\/).

\subsection{Pulse profiles}

1 -- Using the high statistics of all pre-outburst observations, ISGRI and RXTE HEXTE both measure significant and consistent 
pulse profiles up to 150 keV. Namely, a broad single asymmetric pulse, making \psrtar\ the fifth rotation-powered pulsar detected 
at hard X-rays/soft gamma rays ({\slshape {Sect. \ref{sect_tm_pp}, Fig. \ref{isgri_pulse_profile}}}\/).

\noindent
2 -- Comparison of the pulse profiles measured with the PCA, HEXTE and ISGRI between 2.9 and 150 keV shows that the
pulse shape does not change with energy ({\slshape {Sect. \ref{sect_tm_pp}, Figs. \ref{isgri_pulse_profile} \& \ref{profile_collage}}}\/).

\noindent
3 -- The shape of the profile measured with the PCA between 2.9 and 8.3 keV during the magnetar-like outburst, when the soft 
component dominated the emission, exhibits a shape which is fully consistent with the high-statistics pre-outburst pulse 
profile in the same energy band. This is evidence for a stable geometry ({\slshape {Sect. \ref{sect_tm_pmorph}, Fig. 
\ref{profile_comparison}}}\/).

\subsection{Pre-outburst pulsed spectra and pulsed fractions; PWN-DC spectrum}

1 -- The time-averaged spectra of the pre-outburst pulsed emission as measured with the PCA, HEXTE and ISGRI 
between $\sim$ 3 keV and 300 keV are consistent in overlapping energy bands. The best-fit power-law model has
a photon index $\Gamma$ = $1.20\pm 0.01$. The energy fluxes of the pulsed emission in the 2-10 keV and 20-100 keV 
bands are $(2.38\pm0.02)\times 10^{-12}$ erg/cm$^2$s and $(15.2\pm0.14)\times 10^{-12}$ erg/cm$^2$s, respectively. 
Fitting a curved power-law model to just the PCA spectrum (3-30 keV) and to the total combined pulsed spectrum 
(3-300 keV) gives a significantly better fit.
({\slshape {Sect. \ref{sect_pes_ppre}, Figs. \ref{he_spectrum} \& \ref{spc_comparison_rpp_axp}}}\/).

\noindent
2 -- Comparing the measured pulsed flux (only \psrtar\ pulsed) and the total (\psrtar\ pulsed + DC + PWN) measured 
by ISGRI in the band 20-100 keV gives a lower limit to the pulsed fraction of 44\% for the emission from the
compact object. Around 150 keV the measured pulsed and total fluxes are consistent with a pulsed fraction of 
100\%, however the statistical uncertainties are large at these energies.
Comparing the PCA pulsed flux with the total point-source (\psrtar\ pulsed + DC) flux measured 
with Chandra, gives an estimated pulsed fraction of 31\% for the 0.5-10 keV band ({\slshape {Sect. \ref{sect_pes_pf}, 
Fig. \ref{he_spectrum} }}\/).

\noindent
3 -- The above discussed variation of pulsed fraction with energy implies that the high-energy spectrum of the PWN, 
as well as that of the point-source DC emission of \psrtar\ measured by Chandra below 7 keV will strongly bend down 
in the INTEGRAL energy window above 20 keV to drastically lower (${E^2}$F) levels at 100 keV ({\slshape {Sect. \ref{sect_pes_pf}, 
Fig. \ref{he_spectrum}}}\/).

\subsection{Spectral evolution (3-30 keV) during magnetar-like outburst}

1 -- In the first 32 days after the major glitch (time segment A1) the 3-30 keV pulsed spectrum can be represented with two 
power-law models, a soft component with index $\Gamma_{s}$ = $2.96\pm 0.06$ and a hard component with the pre-outburst value 
$\Gamma_{h} = 1.16$. The 2-10 keV (energy) flux increased by a factor $\sim 5$ and the 10-30 keV (energy) flux increased with 
only 35\% ({\slshape {Sect. \ref{sect_pes_spcevol}, Figs. \ref{tp_outburst_gt9kev} \& \ref{excess_ph_flux_lt10}}}\/).

\noindent
2 -- Above 9 keV all spectra during outburst and also the 3-30 keV pre- and post-outburst spectra are consistent with a 
single power-law shape with the same index 1.16 ({\slshape {Sect. \ref{sect_pes_spcevol}, Fig. \ref{spectral_variations}}}\/).

\noindent
3 -- After $\sim$ 120 days the strong soft outburst and the modest enhancement of the hard non-thermal component both vanish ({\slshape {Sect. \ref{sect_pes_spcevol}, Figs. \ref{tp_outburst_gt9kev} \& \ref{excess_ph_flux_lt10}}}\/).


\section{Discussion}

The high-B-field radio-quiet X-ray pulsar \psrtar\ appears to be a unique object. Ever since the discovery of high-energy X-ray and 
gamma-ray emission from rotation-powered (radio) pulsars, they have been known as stable high-energy emitters. \psrtar\ is the first
rotation-powered pulsar exhibiting a magnetar-like radiative outburst together with short $< 0.1$s bursts \citep{kumar08, gavriil08}.  
This schizophrenic behaviour triggered the discussion whether high-B-field rotation-powered pulsars represent a transition between normal,
lower-field pulsars and magnetars. In this discussion we will first address its manifestation as a genuine rotation-powered pulsar and 
whether its high-energy characteristics are commensurate with those of other young rotation-powered pulsars and theoretical scenarios 
discussed in literature for the production of non-thermal emission in pulsar magnetospheres. Then, we will discuss how the characteristics 
which we determined for its magnetar-like outburst point to a likely production site and scenarios in the dipole geometry discussed in 
models for rotation-powered pulsars.

\subsection{\psrtar\ as a rotation-powered pulsar}

Before its brightening and since its discovery in X-rays by \citet{gotthelf2000}, \psrtar\ manifested itself for many years as a
stable pulsar emitting X-rays with energies above 0.5 keV. The total spectrum (Kes 75 plus the \psrtar/PWN system) measured with
INTEGRAL above 20 keV has a photon index $\Gamma = 1.80\pm0.06$ and an X-ray luminosity (20-300 keV) of 
$L_{X,tot}^{20-300}=6.3(3)\times 10^{34}\cdot \eta \cdot d_{10}^2$ erg/s. This luminosity amounts $\sim 9.7\%$ of the available spin-down 
luminosity assuming isotrope emission ($\eta=4\pi$) and a distance of 10 kpc ($d_{10}=1$) and is of the same magnitude, but slightly higher 
than what has been measured for other young pulsars such as the Crab \citep[3.0\%; adopting a distance of 2 kpc and the high-energy model, 
a broken power-law, given in][]{jourdain08} and PSR B1509-58 (5.7\%; adopting a distance of 5.2 kpc) in the same energy band.
Also for energies below 10 keV the total luminosity of \psrtar\/ is similar to that of other young pulsars as discussed by \citet{su2008}. 
Obviously, adoption of a significantly different distance to \psrtar\/ \citep[e.g. 19 kpc by][]{mcbride2008} will lead to different conclusions.

\begin{figure}
  \centering
  \includegraphics[height=9cm,width=9cm]{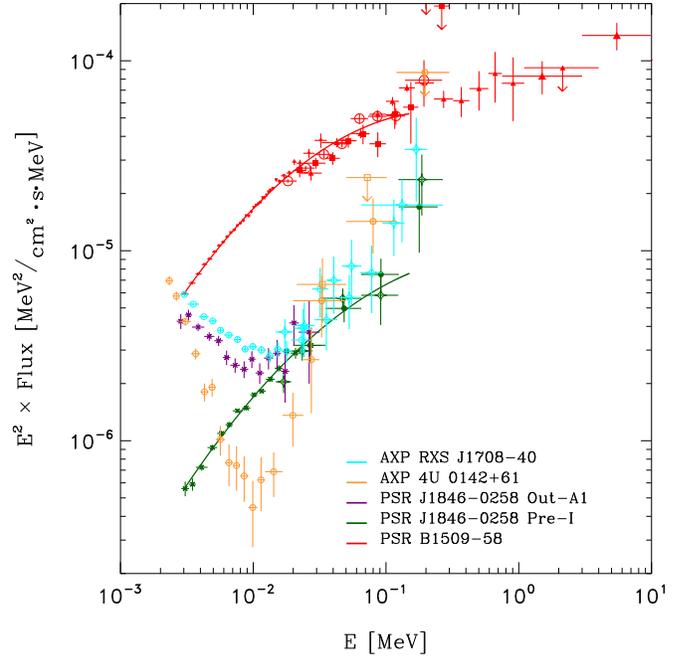}
  \caption{The high-energy pulsed spectra of \psrtar\ from pre-outburst observations with RXTE and INTEGRAL (green data points; green
           curved line) and from observations performed during the early phase of the outburst (A1) from RXTE PCA data (purple 
           data points). For comparison purposes the pulsed spectrum of the young rotation-powered high-B field pulsar PSR B1509-58 is
           superposed (red data points from RXTE PCA and HEXTE, INTEGRAL ISGRI and CGRO BATSE and COMPTEL; red curved line, fit to RXTE 
           PCA and HEXTE data) along with the pulsed spectra (RXTE and INTEGRAL) of AXPs 4U 0142+61 (orange data points) and 
           1RXS J1708-4009 (aqua data points). Error bars are $1\sigma$. All spectra are corrected for interstellar absorption (unabsorbed).}
  \label{spc_comparison_rpp_axp}
\end{figure}

The time-averaged non-thermal pulsed spectrum has a photon index $\Gamma$ = $1.20\pm 0.01$ (3-300 keV, this work) and the X-ray luminosity 
(2-100 keV) of the pulsed emission amounts $L_{X}^{2-100}=1.90(2)\times 10^{34}\cdot \eta \cdot d_{10}^2$ erg/s, which translates in a 
spin-down luminosity efficiency of 0.23\% assuming emission into one steradian ($\eta=1$) and a distance of 10 kpc ($d_{10}=1$).
These values are in the wide range measured for other young rotation powered pulsars emitting at hard X-rays e.g. Crab, PSR 1509-58 
and PSR B0540-69.
For example, \psrtar\ can be compared with the young radio pulsar PSR B1509-58, which also harbours a strong polar surface magnetic field 
with strength of $\sim 1.5 \times 10^{13}$ G, in this case below the quantum critical field strength. Furthermore, PSR B1509-58 
has a similar broad single-pulse profile at soft and hard X-rays, and is detected up to at least 10 MeV \citep{kuiper99, cusumano01}. 
In the 3-300 keV energy band, the shapes of the pulsed non-thermal spectra of both pulsars are similarly curved (see Fig. \ref{spc_comparison_rpp_axp}).

The main difference is that the pulsed flux of \psrtar\ is about a factor of ten lower than the pulsed flux from PSR B1509-58
(see Fig. \ref{spc_comparison_rpp_axp}). Taking the different source distances into account, $5.2\pm1.4$ kpc for PSR B1509-58 and
$\sim 10$ kpc \citep{su2008} for \psrtar, the luminosities differ only by a factor of $\sim 2.7$.
However, the efficiencies to convert spin-down energy into X-rays are (very) comparable, $\epsilon_{1846}/\epsilon_{1509} \sim 0.8$ 
(the spin-down luminosity of PSR B1509-58 amounts $1.77\times 10^{37}$ erg/s, and that of \psrtar\ $8.19\times 10^{36}$ erg/s).
Therefore, \psrtar\ does not appear to be intrinsicly different from other young normal rotation-powered (radio) pulsars.

The scenario responsible for the production of non-thermal X-ray and gamma-ray emission from rotation-powered (radio) pulsars 
has been under debate for decades. There is still no consensus on the origin of the high-energy electrons and/or positrons that are responsible for the high-energy photons, and how and where they are accelerated in the open zone along the magnetic-field-aligned electric field. For many years the discussion was between polar-cap scenarios in which the acceleration was proposed to take place near the neutron star at the magnetic poles \citep{arons1979, daugherty1982}, and outer gap models placing the site of acceleration and emission between the null-charge surface on which the magnetic field becomes perpendicular to the rotation axis and the light cylinder on which the plasma corotates with the speed of light \citep{cheng1986a, cheng1986b, romani1996, hirotani01}. Both approaches encountered problems, e.g. the low-latitude polar-cap emission was difficult to reconcile with double-peaked or wide pulse profiles and the outer-gap scenario
faced difficulties in reproducing e.g. the shape of the Vela pulse profile. 

For the polar-cap approach the solution was found with the slot gap \citep{arons1983}, a narrow bundle of field lines bordering the closed-field region in which acceleration and pair cascades can occur at altitudes up to several stellar radii \citep{muslimov03, muslimov04, dyks04}. Slot gaps are predicted to exist only in the young and/or fast pulsars. In the latest version \citet{harding08} presented results from a 3D model
in which they include emission produced by primary electrons accelerated to high-altitudes in the unscreened electric field of the slot gap, as well as by higher-generation electron-positron pairs cascading from lower altitudes along field lines interior to the slot gap. Curvature, synchrotron and inverse Compton  radiation of both primary electrons and pairs contribute to the total broad-band emission. \citet{harding08} show that the slot-gap model can reasonably well reproduce the Crab-pulsar profiles and (phase-resolved) spectra. The optical to hard X-ray component for this young pulsar is produced by synchrotron radiation of the unaccelerated pairs flowing along the interior field lines of the slot gap to high altitudes. Applying a slot-gap model also for the young \psrtar\, synchrotron radiation of pairs is then responsible for the stable time-averaged pulsed hard-X-ray spectrum (pre- and post-outburst) shown in Fig. \ref{he_spectrum}. The broad single pulse profile of \psrtar\ can be
modelled in this scenario for different viewing directions (angle between observer and rotation axis, and angle between rotation- and magnetic field axis)
than those modelled for the Crab pulsar \citep{harding08}.

For the traditional outer-gap scenario, in which the non-thermal emission is produced above the null-charge 
surface, \citet{zhang00} showed in a 3D model that they could reasonably well reproduce for the two young pulsars 
PSR B1509-58 and PSR B0540-69 both their broad pulse profiles (similar to \psrtar\ ) was well as their 
soft to hard X-ray spectra. However, in the traditional outer-gap model the existence of the outer gap had been hypothesized. \citet{hirotani99} demonstrated in a one-dimensional analysis of a pair-production cascade along the magnetic field line, that the outer gap does exist as solution of the Maxwell and Boltzmann equations. This model was further elaborated in a 2D scenario \citep{takata06, hirotani06, takata07a} and finally the photon propagation and pair production is solved in the full 3D magnetosphere \citep{takata07b, hirotani08}. For these self-consistent solutions the gap extends past the null-charge surface toward the neutron star.  \citet{takata07b} and \citet{hirotani08} showed that the modified outer-gap model can also reproduce the pulse profile and phase-resolved spectra of the Crab. In addition, \citet{takata07b} reproduced the spectrum and broad structured pulse profile for the young PSR B0540-69. Like for the slot-gap scenario,  in the framework of the outer-gap accelerator synchrotron emission is responsible for the emission in the X-ray band. This synchrotron component is dominated by the emissivity of accelerated primary pairs, with a substantial contribution from secondary pairs. Notably, \citet{takata07b} show that hard X-ray synchrotron emission from secondary pairs below the null-charge surface, thus close to the neutron-star surface, contributes up to energies above 1 MeV. At MeV energies this latter synchrotron component even dominates over the synchrotron contribution from primary pairs.

In fact, the geometry of the modified outer-gap model and that of the slot-gap model are now very similar, but the radiation mechanisms are still very different. Both models appear to be reasonably capable of reproducing spectra and pulse profiles of young pulsars. The X-ray spectral and temporal characteristics measured for \psrtar\ before and after the magnetar-like outburst are very similar to those seen for other young pulsars, e.g. PSR 1509-58 and PSR B0540-69, and can also be explained with either model.

\subsection{\psrtar\ exhibiting magnetar-like behaviour}

We were able to study the magnetar-like outburst in detail rendering important directions on what most likely caused the
event, where the initial trigger happened, and what are likely scenarios in the magnetosphere for the production of
the outburst emission above 3 keV.

\subsubsection{Magnetar-like outburst below 10 keV}

Firstly, the evolution of the spectral shape of the pulsed emission over the outburst time interval (see Fig. \ref{spectral_variations}) 
clearly shows that the major glitch of size $\Delta\nu/\nu$  $(2.0-4.4)\times 10^{-6}$ instantly triggered the ejection of a soft component 
with an initial flux increase by a factor $\sim$ 5 for energies 2-10 keV. The pulsed spectrum between 3 and 30 keV can be described for 
the first 125 days (time interval A1, A2 and B in Fig. \ref{outburst_freq_evol} and Table \ref{table:pca_temp_spec}) with the sum of 1) a new 
soft component which appeared to soften over the burst duration (the photon index evolved from $2.96\pm 0.06$ via $3.27\pm 0.27$ to $4.74\pm 0.71$, indicative for cooling), and 2) a non-thermal component with the same photon index of $\sim 1.2$ as the pre-outburst pulsed spectrum, but
initially enhanced in flux by $\sim 35\%$. The new soft component vanished at post-outburst epochs and the non-thermal component returned back to its pre-outburst flux levels. We do not have spectral information on the pulsed emission 
below 3 keV, but \citet{kumar08} and \citet{ng08} show evidence for a black-body (BB) component in the Chandra total point-source spectrum (pulsar
plus DC) seven days after the start of the outburst (during interval A1 in Fig. \ref{outburst_freq_evol}). The estimated temperatures were in the range kT $\sim$ 0.7-1.1 keV with BB radii in the range $\sim$ 0.2-0.6 km, in agreement with the conventional polar-cap radius. These authors did 
not detect a BB component in 2000, when \psrtar\ was in its low persistent state. We do not consider this non-detection as being due to low
statistics, because the high-energy X-ray and gamma-ray spectra of young pulsars are known to be dominated by non-thermal
emission and do not exhibit a BB enhancement. The latter becomes apparent for middle-aged pulsars like Vela and older pulsars.
An exception is PSR J1119-6127, which is also a young high-magnetic-field pulsar with spin properties very similar to
those of \psrtar, but with different X-ray properties. \citet{safi-harb08} show that the spectrum of PSR J1119-6127 has a thermal component which can be fitted by either a BB model with a temperature kT $\sim$ 0.21 keV, or a neutron star atmospheric model with temperature kT $\sim$ 0.14 keV. Both temperatures are significantly lower than that measured during outburst for \psrtar. Therefore, we conclude that the BB component of \psrtar\ during outburst is like the soft component measured with RXTE initiated by the major timing glitch and cools down in $\sim$ 120 days. Then, the measured soft transient
component of the pulsed emission between 3 and 10 keV {in excess to the underlying hard non-thermal component} can be due to resonant cyclotron scattering (RCS) thereof, as has been proposed for e.g. magnetars for their persistent and transient emissions \citep[][]{thompson02,beloborodov07}. For example, \citet{rea08} applied this model to the total X-ray spectra of 10 magnetars, including canonical and transient AXPs and SGRs. 
Fig. \ref{spc_comparison_rpp_axp} shows also the pulsed high-energy spectra of AXPs 4U 0142+61 and 1RXS J170849-400910, again measured with RXTE 
and INTEGRAL \citep{denhartog08a, denhartog08b}, to be compared with the pulsed spectrum of \psrtar\ directly after the major glitch. The similarity in shape is evident. The pulsed spectra between 3 and 30 keV of 4U 0142+61 and 1RXS J170849-400910 can also be represented by the sum of two, a soft and a hard, power-law components. The soft component can plausibly be explained with a resonant cyclotron scattering origin. The hard non-thermal power-law spectra of 4U 0142+61 and 1RXS J170849-400910 have been measured with INTEGRAL up to $\sim$ 200 keV and could have a similar origin in 
the outer magnetosphere as the hard non-thermal X-ray spectra measured for young radio pulsars, as has been discussed in \citet{denhartog08b}.
Furthermore, and most interestingly, we found that during the outburst the pulse profile in this energy range below $\sim$ 9 keV, while completely dominated by the soft flux enhancement, is identical to the pulse profile before and after the outburst when the emission was purely non-thermal (Fig. \ref{profile_comparison}). This means that the soft enhancement is produced in the same viewing direction through the pulsar magnetosphere. Unfortunately, there is no measurement of the profile shape during outburst below 3 keV, where the soft component dominates.

\subsubsection{Flux enhancement above 10 keV}
In Section \ref{sect_pes_spcevol} we showed that above 10 keV, the evolution over the outburst period of the pulsed spectrum differs from that
below 10 keV. Namely, the pulsed flux (10-30 keV) increased after the glitch by only $\sim$ 35\%, returning to the pre-outburst level in $\sim$ 120 days together with the quenching of the new soft component. The pulsed spectrum (10-30 keV) remained hard during the outburst, consistent with the pre- and post-outburst index value of $\sim$ 1.2. Unfortunately, INTEGRAL did not observe \psrtar\ when the glitch triggered the outburst. Nevertheless, INTEGRAL observations during the tail of the outburst covered also PCA observation-time window B (Table \ref{table:pca_temp_spec} and see overlap in Fig. \ref{outburst_freq_evol}), but the statistics were insufficient for extracting the pulsed signal. They revealed, however, an increase of the total flux (20-100 keV) with ($52 \pm 23$)\% and no change in spectral index. This is a indication that the hard X-ray spectrum from \psrtar\ changed only in normalization and not in shape over the broad 10-100 keV energy window,
suggesting that the production mechanism of the non-thermal emission in the pulsar magnetosphere did not change during the outburst.
Furthermore, the shape of the pulse profile above 10 keV, as measured with the PCA, also did not change, suggesting that the geometry of the production site of the non-thermal emission in the pulsar magnetosphere did not alter either. 

\subsubsection{Production scenarios during magnetar-like outburst}

Summarizing, our results strongly suggest that during the outburst, there are two pulsed components: 1) a transient soft component which dominates the pulsed emission below 10 keV, and exhibits the same pulse profile as the pre-outburst non-thermal emission; 2) a non-thermal hard
component with the same characteristics (spectral shape and pulse profile) as the pre- and post-outburst pulsed emission, but
with an $\sim$ 35-\% enhanced flux averaged over 32 days after the major glitch. Apparently, the geometry of the magnetic field in the magnetosphere did not vary, i.e. the assumed pulsar dipole field did not change during the magnetar-like outburst.

During the outburst, the pulsed spectrum of \psrtar\ looks very much like the spectra of the persistent emissions from AXPs,
as we showed in Fig. \ref{spc_comparison_rpp_axp} for AXP 4U 0142+61 and 1RXS J170849-400910. For AXPs strong variability and outbursts were detected
for energies below $\sim$ 10 keV, while the hard non-thermal power-law emissions measured with INTEGRAL up to $\sim$ 200 keV remained stable within $\sim$ 20\% on time scales of 0.5-1 year \citep{denhartog08a, denhartog08b}. Attempts to see correlated flux variability between the soft emission below 10 keV and the hard emission above 10 keV on shorter time scales were not succesful due to the poor statistics in all observations above 10 keV with RXTE HEXTE and INTEGRAL ISGRI. Therefore, similar correlated flux variability for AXPs can not be excluded.
The abrupt flux enhancements of AXPs (and SGRs) below $\sim$ 10 keV have been interpreted as a sudden release of energy, either from below the crust or above. Apparently, we witness here for the first time a similar event in the case of a high-B-field rotation-powered pulsar, and the following scenarios emerge, which can explain the soft-hard-correlated flux variability. Namely, sufficient stress has been built up in the crust to cause cracking, resulting in a major glitch, releasing at the same time a significant amount of energy, generating the BB-component measured with Chandra in the total emission during the outburst.  Resonant cyclotron upscattering thereof subsequently generates the decaying / cooling soft pulsed component measured below 10 keV at the phase of the pulse of the non-thermal component. This suggests that the event causing the abrupt rise, occured near the polar cap (no change in pulse profile).
In the two models discussed above for rotation-powered pulsars, slot-gap and out-gap models, the generation of the pulsed non-thermal emission starts above the polar cap near the neutron star surface. The modest flux increase of the non-thermal emission can then be explained in both model scenarios: 1) In the case of the slot-gap model, an abrupt increase in the number of electron-positron pairs above the polar cap, will temporaly increase the number of unaccelerated pairs flowing along the interior field lines of the slot gap to high altitudes producing the synchrotron radiation responsible for the non-thermal hard X-rays. This provides a natural explanation for the stable pulse profile, because the geometry did not change. 2) For the extended-outer-gap model we are similarly dealing with enhanced synchrotron emission, namely from the abrupt enhancement of secondary pairs created above the
polar cap below the null-charge surface, which are accelerated in the extended gap and contribute up to energies above 1 MeV.

The above scenarios would imply that \psrtar\ did not transform from a rotation-powered pulsar into a magnetar, rather that rotation-powered pulsars with a strong B field can occasionally experience major glitches, releasing significant amounts of energy from below the crusts, following subsequently one of the above sketched scenarios. With the increase of the B-field strengths going from rotation-powered pulsars to AXPs and SGRs,
the frequency of such events increases.

\begin{acknowledgements}
We thank Fotis Gavriil for providing the PCA pulsed flux measurements of \psrtar\/ as shown in Fig. 2 of \citet{gavriil08}.
This research has made use of data obtained from the High Energy Astrophysics 
Science Archive Research Center (HEASARC), provided by NASA's Goddard Space Flight Center,
and of data obtained through the INTEGRAL Science Data Centre (ISDC), Versoix, Switzerland.
INTEGRAL is an ESA project with instruments and science data centre funded by ESA member 
states (especially the PI countries: Denmark, France, Germany, Italy, Switzerland, Spain), 
Czech Republic and Poland, and with the participation of Russia and the USA.
We have extensively used NASA's Astrophysics Data System (ADS).
\end{acknowledgements}

\end{document}